\documentclass[aps,pra,twocolumn]{revtex4}
\usepackage[a4paper,top=1.8cm,bottom=1.8cm,left=1.8cm,right=1.8cm]{geometry}
\usepackage{graphicx}
\usepackage{amsmath,amssymb}
\usepackage{dcolumn}
\usepackage{epsfig}
\allowdisplaybreaks
\newcommand{\ket}[1]{\ensuremath{\left|#1\right>}}
\newcommand{\bra}[1]{\ensuremath{\left<#1\right|}}
\newcommand{\mf}[1]{\boldsymbol{#1}}
\newcommand{\mc}[1]{\ensuremath{\mathcal{#1}}}

\def\email{\footnote{benhurcris@yahoo.com}}
\def\correspondingauthor{\footnote{Corresponding author: rarun@cutn.ac.in}}
\begin{document}
\title{Fluorescence control through vacuum induced coherences}
\author{H. B. Crispin\email{}}
\author{R. Arun\correspondingauthor{}}
\affiliation{Department of Physics, School of Basic $\&$ Applied Sciences, Central University of
	Tamilnadu, Thiruvarur 610005, Tamilnadu, India.}
\date{\today}
\begin{abstract}
The resonance fluorescence of a four-level atom in $J=1/2$ to $J=1/2$ transition driven by two coherent fields is studied. We find
that the incoherent fluorescence spectrum shows a direct indication of vacuum-induced coherence in the atomic system. We show that such
coherence manifests itself via an enhancement or suppression of the spectral peaks in the $\pi$-polarized fluorescence. The effect of
the relative phase of the driving fields on the spectral features is also investigated. We show that phase-dependent enhancement or
suppression of the fluorescence peaks appears in the incoherent spectrum emitted along the $\sigma$ transitions. It is found that this
phase dependence occurs because of the polarization-detection scheme employed for the observation of the fluorescence light. We present an
analytical explanation, based on dressed-states of the atom-field system, to interpret the numerical results.
\end{abstract}
\maketitle
\newpage
\section{\label{sec:intro}INTRODUCTION}
\vspace{-0.5em}
In the past few decades, considerable effort has been devoted to the study of coherence and interference effects arising from the spontaneous
emission of atoms and the subject has been reviewed in detail by Ficek and Swain \cite{ficek1,ficek2}. It is well understood how the spontaneous
decay of closely lying energy states coupled by common vacuum modes leads to a new type of coherence between the states
\cite{car2,java,zhu1,zho,gong,bor,ever1,gao,zhu2,paspa,li1,li2,hou,arun1,yan,si,arun2}. Even in the absence of an external driving field, a coherence
can be induced between the excited levels of an atom due to the vacuum field. This type of coherence is known as vacuum induced coherence (VIC)
in the literature \cite{ficek2}. The effects of VIC depend on the level structure of the atom and the dipole moments of the atomic transitions
 involved in the dynamics. Early studies focussed on three-level atoms ($V$- or $\Lambda$-type configurations) and found modifications of the
 fluorescence, absorption, and dispersion properties of the atomic medium due to VIC, such as quenching of fluorescence \cite{car2}, disappearance
 of coherent population trapping state \cite{java}, spectral line narrowing and dark lines in the spectrum \cite{zhu1}, ultrasharp spectral
 lines \cite{zho}, probe light amplification with and without population inversion \cite{gong}, phase control of pulse propagation \cite{bor}
 and population dynamics \cite{ever1}, and enhancement of squeezing \cite{gao}. The role of VIC has been explored in four-level atomic systems
 as well \cite{zhu2,paspa,li1,li2,hou,arun1,yan,si,arun2}. Many interesting features such as spectral line elimination and spontaneous emission
 cancellation \cite{zhu2}, phase control of spontaneous emission \cite{paspa}, fluorescence suppression and line narrowing \cite{li1},
 interference-assisted squeezing in resonance fluorescence \cite{li2}, inhibition of two-photon transparency \cite{hou}, interference in cascade
 spontaneous emission \cite{arun1}, enhancement of self-Kerr nonlinearity \cite{yan} and non-linear dispersion \cite{si}, and superluminal
 light propagation \cite{arun2} have been reported.

In all these publications, the VIC effects are studied based on the assumption that the dipole transition moments in atoms are non-orthogonal.
This condition, being a stringent requirement for the existence of VIC in atoms, is difficult to achieve in practice. To overcome this difficulty,
 many alternative methods have been proposed to bypass the requirement of non-orthogonal dipole transitions \cite{bypass}. In a remarkable paper,
 Kiffner {\it et al.} proposed yet another scheme to realize VIC in atomic systems \cite{kif1}. They considered the resonance fluorescence
 from a $J=1/2$ to $J=1/2$ transition which is driven by a linearly polarized field acting on the $\pi$ transitions. It was shown that the system
 exhibits VIC effects even though the $\pi$ transitions do not share  common initial and final states \cite{kif1,kif2}. The advantage of this
 configuration is that its level structure is found in ${}^{198}\text{Hg}^+$ ions \cite{polder,hgion} making it a suitable candidate for verifying VIC
 features. Due to the realistic nature of this scheme, a number of other studies on the consequences of VIC in this system has been done.
 Das {\it et al.} reported that the inclusion of VIC results in a larger value of the second order correlation function \cite{gs4}. Further,
 studies of the squeezing spectrum \cite{tan} and interaction between two dipole-dipole interacting four-level atoms of this scheme \cite{ever2}
  have shown that the VIC does play a prominent role in modifying the fluorescence properties of the system.

\begin{figure}[b]
\begin{center}
\includegraphics[width=8cm,height=5cm]{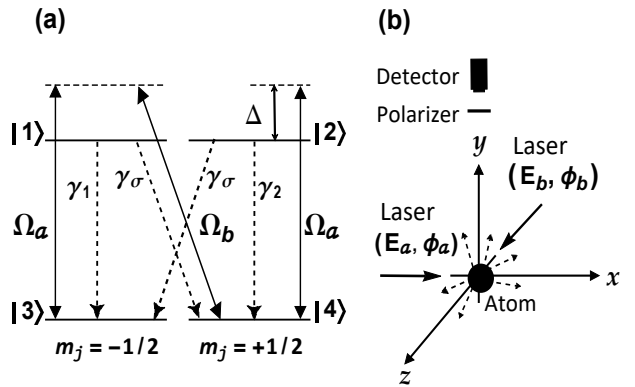}
\caption{(a) The level scheme of a four-level atom with $J=1/2$ to $J=1/2$ transitions driven by coherent fields. The transitions
$\ket{1}\leftrightarrow\ket{3}$ and $\ket{2}\leftrightarrow\ket{4}$ are driven by a linearly polarized field while a $\sigma^{-}$-polarized
field induces the transitions $\ket{1}\leftrightarrow\ket{4}$ in the atom. (b) The arrangement for laser fields driving the atom and the
detection of the fluorescence spectrum.}
\end{center}
\end{figure}

In this paper, we consider a four-level atom with $J=1/2$ to $J=1/2$ transition as in earlier publications \cite{kif1,kif2,polder,hgion,gs4,tan,ever2}.
The atom has excited and ground levels that are both doubly degenerate, as shown in Fig. 1(a). Since the spontaneous decays along the $\pi$-transition
channels $\ket{1} \rightarrow \ket{3}$ and $\ket{2} \rightarrow \ket{4}$ occur via common vacuum modes, VIC exists in this system.
In the previous studies, the fluorescence properties of the atom were investigated considering only a linearly polarized light driving the
$\pi$ transitions \cite{kif1,kif2,gs4,tan,ever2}. In the present work, we extend this analysis to include an additional $\sigma^{-}$-polarized
light driving the atom. We study the effects of VIC in the resonance fluorescence from the driven system. We will show that many interesting
features arise in the fluorescence spectrum such as appearance of a narrow central peak, splitting of sidebands, enhancement and suppression of
fluorescence peaks, and phase-dependent spectral features.

Our paper is arranged as follows. In Sec. \ref{sec:atom}, we discuss the Hamiltonian of our system and present the density matrix equations.
In Sec. \ref{sec:stdypop}, we examine the atomic population distribution in steady state and discuss the role of additional $\sigma^{-}$ polarized
light driving the system. The expressions for the incoherent fluorescence spectra of the $\pi$ and $\sigma$ transitions are then derived in
Sec. \ref{sec:spect}. In Sec. \ref{sec:num}, we present the numerical results of the fluorescence spectrum and analyze the new features using a
dressed-state description. Finally, the main results are summarized in Sec. \ref{sec:concl}.

\section{\label{sec:atom}ATOMIC SYSTEM AND DENSITY MATRIX EQUATIONS}
\vspace{-0.5em}
The atomic model under consideration has two degenerate excited and ground levels which can be realized with $J=1/2$ to $J=1/2$ transition [see Fig. 1(a)].
Each of the excited atomic states $\left(\ket{1} \text{and} \ket{2}\right)$ decays via spontaneous emission to both the ground states
$\left(\ket{3} \text{and} \ket{4}\right)$. The transitions $\ket{1}\leftrightarrow\ket{3}$ and $\ket{2}\leftrightarrow\ket{4}$ are referred to as the
$\pi$ transitions, whereas the cross transitions $\ket{1}\leftrightarrow\ket{4}$ and $\ket{2}\leftrightarrow\ket{3}$ in the atom are designated
as $\sigma$ transitions. Direct transitions between the excited states $\left(\ket{1}\leftrightarrow\ket{2}\right)$ as well as between the ground
states $\left(\ket{3}\leftrightarrow\ket{4}\right)$ are assumed to be dipole forbidden. The $\pi$ transitions have antiparallel dipole moments and
can couple with light linearly polarized along the z-direction $(\mf{e}_{z})$. The $\sigma$ transitions $\ket{1}\leftrightarrow\ket{4}$ and
$\ket{2}\leftrightarrow\ket{3}$ can couple to $\sigma^{-}$ and $\sigma^{+}$ polarized light, respectively. The transition dipole moments can be
calculated from the matrix elements of the electric-dipole moment operator $\mf{\hat{d}}$. Using the Wigner-Eckart theorem \cite{sak}, they are
obtained as
\begin{eqnarray}
\mf{d}_{1}&=&\bra{1}\mf{\hat{d}}\ket{3}=\frac{-1}{\sqrt{3}}\,\mc{D}\,\mf{e}_{z},~~ \mf{d}_{2}=\bra{2}\mf{\hat{d}}\ket{4}=-\mf{d}_{1},\nonumber\\
\mf{d}_{3}&=&\bra{2}\mf{\hat{d}}\ket{3}=\sqrt{\frac{2}{3}}\,\mc{D}\,\mf{\epsilon^{(-)}},~\mf{d}_{4} = \bra{1}\mf{\hat{d}}\ket{4}=\mf{d}_{3}^{*},\label{matel}
\end{eqnarray}
where $\mf{\epsilon}^{(-)}=\left(\mf{e}_{x} - i\,\mf{e}_{y}\right)/\sqrt{2}$ denotes the circular polarization vector and \mc{D} is the reduced
dipole matrix element.

We are interested in the situation in which two coherent fields of equal frequencies drive the atom. The coherent fields propagate in perpendicular
directions and interact with the atomic system as shown in Fig. 1(b). The transitions $\ket{1}\leftrightarrow\ket{3}$ and
$\ket{2}\leftrightarrow\ket{4}$ are coupled  by a linearly polarized field (amplitude $E_{a}$, phase $\phi_{a}$, polarization $\mf{e}_{z}$) travelling
in the x-direction. A circularly polarized field (amplitude $E_{b}$, phase $\phi_{b}$, polarization $\mf{\epsilon}^{(-)}$) propagating along the
z-direction is set to drive the transitions $\ket{1}\leftrightarrow\ket{4}$ in the atom. The Rabi frequency of the linearly (circularly) polarized
field driving the atom is denoted as $\Omega_{a}$ $(\Omega_{b})$. The Hamiltonian for this atom-field system is given in the dipole and rotating-wave
approximations to be
\begin{flalign}
H=&~\hbar\omega_{o}(A_{11}+A_{22})+\hbar[\Omega_{a}(A_{13}-A_{24})e^{-i(\omega_{l}t+\phi_{a})}&\nonumber\\
&-\Omega_{b}A_{14}e^{-i(\omega_{l}t+\phi_{b})}+h.c.],&
\end{flalign}
where $\omega_{o}=\omega_{13}=\omega_{24}$ is the atomic transition frequency, $\omega_{l}$ is the frequency of both the applied fields,
$A_{mn}=\ket{m}\bra{n}$ denotes the atomic transition operators for $m\neq n$ and population operators for $m=n$, and the Rabi frequencies are
given by $\Omega_{a}=\mc{D}\,E_{a}/(\sqrt{3}\hbar)$ and $\Omega_{b}=\sqrt{2}\mc{D}\,E_{b}/(\sqrt{3}\hbar)$.

The time evolution of the system is studied using the density matrix formalism. The spontaneous emissions in the atom are included via master
equation approach.  The master equation for the reduced density operator $\tilde{\rho}$ of the atomic system in the Schr\"{o}dinger picture is given by
\begin{equation}
\frac{d\tilde{\rho}}{dt}=-\frac{i}{\hbar}[H,\tilde{\rho}]+\mc{L}\tilde{\rho}.
\end{equation}
Here the Liouville operator $\mc{L}\tilde{\rho}$ describes the damping terms due to spontaneous decay processes. We choose the following unitary
transformation to remove the fast-oscillating as well as phase-dependent exponential terms in the interaction
\begin{equation}
\text{U}= \exp\{i[\omega_{l}t+\phi_{a}](A_{11}+A_{22})+i[\phi_{a}-\phi_{b}](A_{22}+A_{44})\}.\nonumber
\end{equation}
The transformed master equation for the density operator $\rho=\text{U}\,\tilde{\rho}\,\text{U}^{\dagger}$ in the
interaction picture becomes
\begin{equation}
\frac{d\rho}{dt}=-\frac{i}{\hbar}[H_{I},\rho]+\mc{L}\rho.\label{masint}
\end{equation}
In Eq. (\ref{masint}), the Hamiltonian of the atom-field system is given by
\begin{flalign}
H_{I}=&-\hbar\Delta(A_{11}+A_{22})&\nonumber\\&+\hbar(\Omega_{a}(A_{13}-A_{24})-\Omega_{b}A_{14}+h.c.),\label{hamil} &
\end{flalign}
where $\Delta=\omega_{l}-\omega_{o}$ is the detuning of the applied fields from the atomic resonance frequency. The damping term $\mc{L}\rho$ is given by
\begin{align}
\mc{L}\rho=&-\frac{\gamma_{1}}{2}(\rho A_{11}+A_{11} \rho-2A_{31} \rho A_{13})\nonumber\\
&-\frac{\gamma_{2}}{2}(\rho A_{22}+A_{22} \rho-2 A_{42} \rho A_{24})\nonumber\\
&-\frac{\gamma_{\sigma}}{2}(\rho A_{11}+A_{11} \rho-2 A_{41} \rho A_{14})\nonumber\\
&-\frac{\gamma_{\sigma}}{2}(\rho A_{22}+A_{22} \rho-2 A_{32} \rho A_{23})\nonumber\\
&+\gamma_{12}(A_{42} \rho A_{13}+A_{31} \rho A_{24}), \label{lrho}
\end{align}
where $\gamma_{1}=\gamma_{2}=\gamma/3$ and $\gamma_{\sigma}=2\gamma/3$ are the decay rates of the $\pi$ and $\sigma$ transitions, respectively
[see Fig. 1(a)]. Note that $\gamma = \gamma_{1}+\gamma_{\sigma} = \gamma_{2}+\gamma_{\sigma}$ gives the total decay rate of each of the excited
atomic states. The cross-damping term $\gamma_{12}$ in Eq. (\ref{lrho}) is responsible for VIC effects in the atom and arises because the spontaneous
decays along the transitions $\ket{1} \rightarrow \ket{3}$ and $\ket{2} \rightarrow \ket{4}$ occur via common vacuum modes. It is given by
$\gamma_{12} = (\mf{d_{1}}\cdot\mf{d_{2}^*}/|\mf{d_{1}}||\mf{d_{2}}|)\sqrt{\gamma_{1}\gamma_{2}}= -\sqrt{\gamma_{1}\gamma_{2}}$, where the minus sign
comes from the anti-parallel dipole moments $\mf{d_{1}}$ and $\mf{d_{2}}$.
If the $\gamma_{12}$-term is ignored $(\gamma_{12} = 0)$, then there is no VIC effect in spontaneous emission.

To study the dynamical behavior of the driven atom, we use the master equation~(\ref{masint}) in the interaction picture.  The equations of motion of
the density matrix elements in the atomic-state basis then take the form
\begin{flalign}
\dot{\rho}_{11}=&-(\gamma_{1}+\gamma_{\sigma})\rho_{11}+i\Omega_{a}(\rho_{13}-\rho_{31})-i\Omega_{b}(\rho_{14}-\rho_{41}),&
\label{rho1}
\end{flalign}
\begin{flalign}
\dot{\rho}_{33}=&~\gamma_{1}\rho_{11}+\gamma_{\sigma}\rho_{22}-i\Omega_{a}(\rho_{13}-\rho_{31}),&
\label{rho2}
\end{flalign}
\begin{flalign}
\dot{\rho}_{44}=&~\gamma_{\sigma}\rho_{11}+\gamma_{2}\rho_{22}+i\Omega_{a}(\rho_{24}-\rho_{42})+i\Omega_{b}(\rho_{14}-\rho_{41}),&
\label{rho3}
\end{flalign}
\begin{flalign}
\dot{\rho}_{12}=&~\left(\!-\frac{(\gamma_{1}+\gamma_{2})}{2}-\gamma_{\sigma}\!\right)\rho_{12}-i\Omega_{a}(\rho_{32}+\rho_{14})&\nonumber\\
&+i\Omega_{b}\rho_{42},&
\label{rho4}
\end{flalign}
\begin{flalign}
\dot{\rho}_{13}=&~\left(\!-\frac{(\gamma_{1}+\gamma_{\sigma})}{2}+i\Delta\!\right)\rho_{13}+i\Omega_{a}(\rho_{11}-\rho_{33})&\nonumber\\
&+i\Omega_{b}\rho_{43},&
\label{rho5}
\end{flalign}
\begin{flalign}
\dot{\rho}_{23}=&~\left(\!-\frac{(\gamma_{2}+\gamma_{\sigma})}{2}+i\Delta \!\right)\rho_{23}+i\Omega_{a}(\rho_{21}+\rho_{43}),&
\label{rho6}
\end{flalign}
\begin{flalign}
\dot{\rho}_{14}=&~\left(\!-\frac{(\gamma_{1}+\gamma_{\sigma})}{2}+i\Delta \!\right)\rho_{14}-i\Omega_{a}(\rho_{12}+\rho_{34})&\nonumber\\
&-i\Omega_{b}(\rho_{11}-\rho_{44}),&
\label{rho7}
\end{flalign}
\begin{flalign}
\dot{\rho}_{24}=&~\left(\!-\frac{(\gamma_{2}+\gamma_{\sigma})}{2}+i\Delta \!\right)\rho_{24}-i\Omega_{a}(\rho_{22}-\rho_{44})&\nonumber\\
&-i\Omega_{b}\rho_{21},&
\label{rho8}
\end{flalign}
\begin{flalign}
\dot{\rho}_{34}=&~\gamma_{12}\rho_{12}-i\Omega_{a}(\rho_{32}+\rho_{14})-i\Omega_{b}\rho_{31}.&
\label{rho9}
\end{flalign}
In writing Eqs. (\ref{rho1})-(\ref{rho9}), we have assumed that the constraint $\rho_{11}+\rho_{22}+\rho_{33}+\rho_{44}=1$ is satisfied at all
times. Note that the VIC term $(\gamma_{12})$ couples the ground-state and excited-state coherences as seen in Eq. (\ref{rho9}). This term plays
a crucially important role in modifying the fluorescence properties of the atom \cite{kif1,kif2}.
\section{\label{sec:stdypop}STEADY-STATE POPULATIONS}
\vspace{-1em}
We first study the population distribution in the atomic levels by solving the density matrix equations. For this purpose, we rewrite the equations
(\ref{rho1})-~(\ref{rho9}) in a compact form as
\begin{equation}
\frac{d}{dt}\mf{\hat{\psi}(t)}=\hat{M}\,\mf{\hat{\psi}(t)}+\hat{C},\label{stdyeq}
\end{equation}
where $\mf{\hat{\psi}}$ is a column vector of density matrix elements
\begin{align}
\mf{\hat{\psi}} = &\left(\langle A_{11}\rangle,\langle A_{33}\rangle,\langle A_{44}\rangle,\langle A_{12}\rangle,\langle A_{21}\rangle,\langle A_{13}\rangle,
\langle A_{31}\rangle,\right.&\nonumber\\&\left.\langle A_{23}\rangle,\langle A_{32}\rangle,\langle A_{14}\rangle,\langle A_{41}\rangle,\langle A_{24}\rangle,
\langle A_{42}\rangle,\langle A_{34}\rangle,\langle A_{43}\rangle\right)^{T}\!\!\!,\label{psicol}
\end{align}
with $\langle A_{ij}\rangle=\rho_{ji}$ and $\hat{C}$ is also a $15\times1$ column vector with non-zero elements
$\hat{C}_{2}=\gamma_{\sigma},\hat{C}_{3}=\gamma_{2},\hat{C}_{12}=i\Omega_{a},\hat{C}_{13}=-i\Omega_{a}$. The inhomogeneous term $\hat{C}$ arises because of
the elimination of the population $\rho_{22}$ in Eqs. (\ref{rho1})-~(\ref{rho9}) using the trace condition $(\hbox{Tr}\rho=1)$. In Eq. (\ref{stdyeq}),
$\hat{M}$ is a $15\times15$ matrix whose elements are independent of the density matrix elements and can be obtained explicitly by using $\mf{\hat{\psi}}$
in Eqs. (\ref{rho1})-~(\ref{rho9}).

The stationary solution of Eq. (\ref{stdyeq}) is obtained by setting $d\mf{\hat{\psi}}/dt = 0$ in the long-time limit. Solving the resulting equation
$\mf{\hat{\psi}}(\!\infty\!)=-\hat{M}^{-1}\,\hat{C}$ gives the steady-state values of the density matrix elements as
\begin{flalign}
\rho_{11} = \rho_{22}
&~=\frac{4\Omega_{a}^{4}}{2\Omega_{a}^{2}(\gamma^{2}+4\Delta^{2}+8\Omega_{a}^{2})+\Omega_{b}^{2}(\gamma^{2}+4\Delta^{2})},&\nonumber
\end{flalign}
\begin{equation}
\rho_{33}=\frac{4\Omega_{a}^{4}+(\Omega_{a}^{2}+\Omega_{b}^{2})(\gamma^{2}+4\Delta^{2})}{2\Omega_{a}^{2}(\gamma^{2}+4\Delta^{2}+8\Omega_{a}^{2})
+\Omega_{b}^{2}(\gamma^{2}+4\Delta^{2})},\nonumber
\end{equation}
\begin{equation}
\rho_{44}=\frac{\Omega_{a}^{2}(\gamma^{2}+4\Delta^{2}+4\Omega_{a}^{2})}{2\Omega_{a}^{2}(\gamma^{2}+4\Delta^{2}+8\Omega_{a}^{2})+\Omega_{b}^{2}
(\gamma^{2}+4\Delta^{2})},
\nonumber
\end{equation}
\begin{flalign}
\rho_{13}= - \rho_{24} &~=\frac{4\Omega_{a}^{3}(\Delta-i\gamma/2)}{2\Omega_{a}^{2}(\gamma^{2}+4\Delta^{2}+8\Omega_{a}^{2})+\Omega_{b}^{2}
(\gamma^{2}+4\Delta^{2})},&\nonumber
\end{flalign}
\begin{equation}
\rho_{23}=\frac{-4\Omega_{a}^{2}\Omega_{b}(\Delta-i\gamma/2)}{2\Omega_{a}^{2}(\gamma^{2}+4\Delta^{2}+8\Omega_{a}^{2})+\Omega_{b}^{2}(
\gamma^{2}+4\Delta^{2})}, \label{sted}
\end{equation}
\begin{equation}
\rho_{34}=\frac{\Omega_{a}\Omega_{b}(\gamma^{2}+4\Delta^{2})}{2\Omega_{a}^{2}(\gamma^{2}+4\Delta^{2}+8\Omega_{a}^{2})+\Omega_{b}^{2}
(\gamma^{2}+4\Delta^{2})},\nonumber
\end{equation}
\begin{equation}
\rho_{12} = \rho_{14} = 0. \nonumber
\end{equation}
As seen in Eqs. (\ref{sted}), the steady-state results for the populations and coherences are independent of the VIC parameter $(\gamma_{12})$ and
the phases $(\phi_{a},\phi_{b})$ of the applied fields. For $\Omega_{b} = 0$, the results (\ref{sted}) become identical to those of Kiffner
{\it et al.} \cite{kif1,kif2}. The coherences $\rho_{23}$ and $\rho_{34}$ are non-zero only when the additional $\sigma^{-}$-polarized field
$(\Omega_{b}\neq 0)$ drives the atom. Also, it is easy to see that $\rho_{33}>\rho_{44}$ for $\Omega_{b}\neq 0$.  An important point is that
the two-photon coherence $\rho_{34}$ is non-zero even though the one-photon coherence $\rho_{14}$ is zero. This implies that the population in the
state $|4\rangle$ can be pumped into the state $|3\rangle$ by two-photon transitions $|4\rangle \rightarrow |1\rangle \rightarrow |3\rangle$ under
the action of both the applied fields $(\Omega_{a},\Omega_{b}\neq 0)$. One thus expects the steady-state population in the ground state $|3\rangle$
to be greater than that of the state $|4\rangle$ unlike the results of Kiffner {\it et al.} \cite{kif1,kif2} as mentioned above. This feature is
illustrated in Fig. 2, where we compare the population distribution in the atom for the cases with and without the additional field.
\begin{figure}[t]
	\begin{center}
		\includegraphics[width=6.2cm,height=4.2cm]{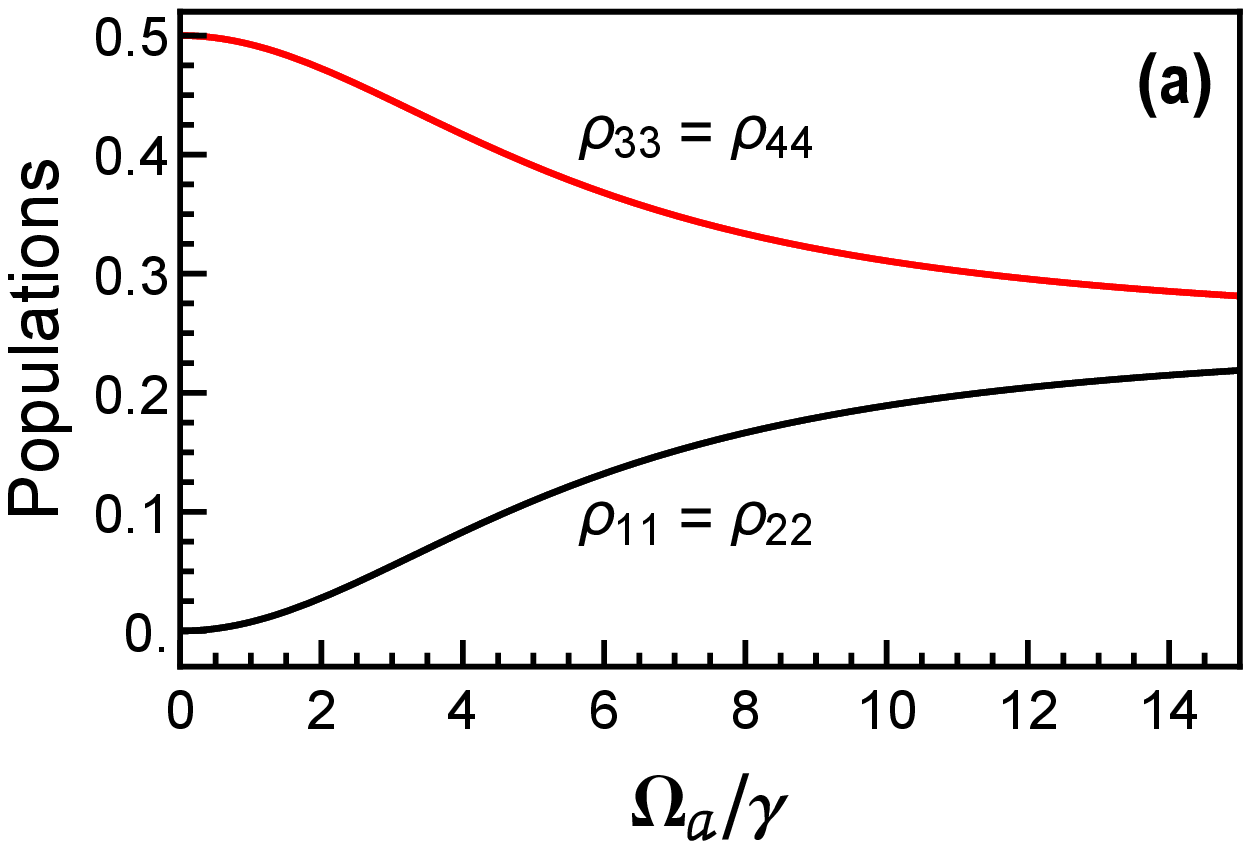}
	\end{center}
	\begin{center}
		\includegraphics[width=6.2cm,height=4.2cm]{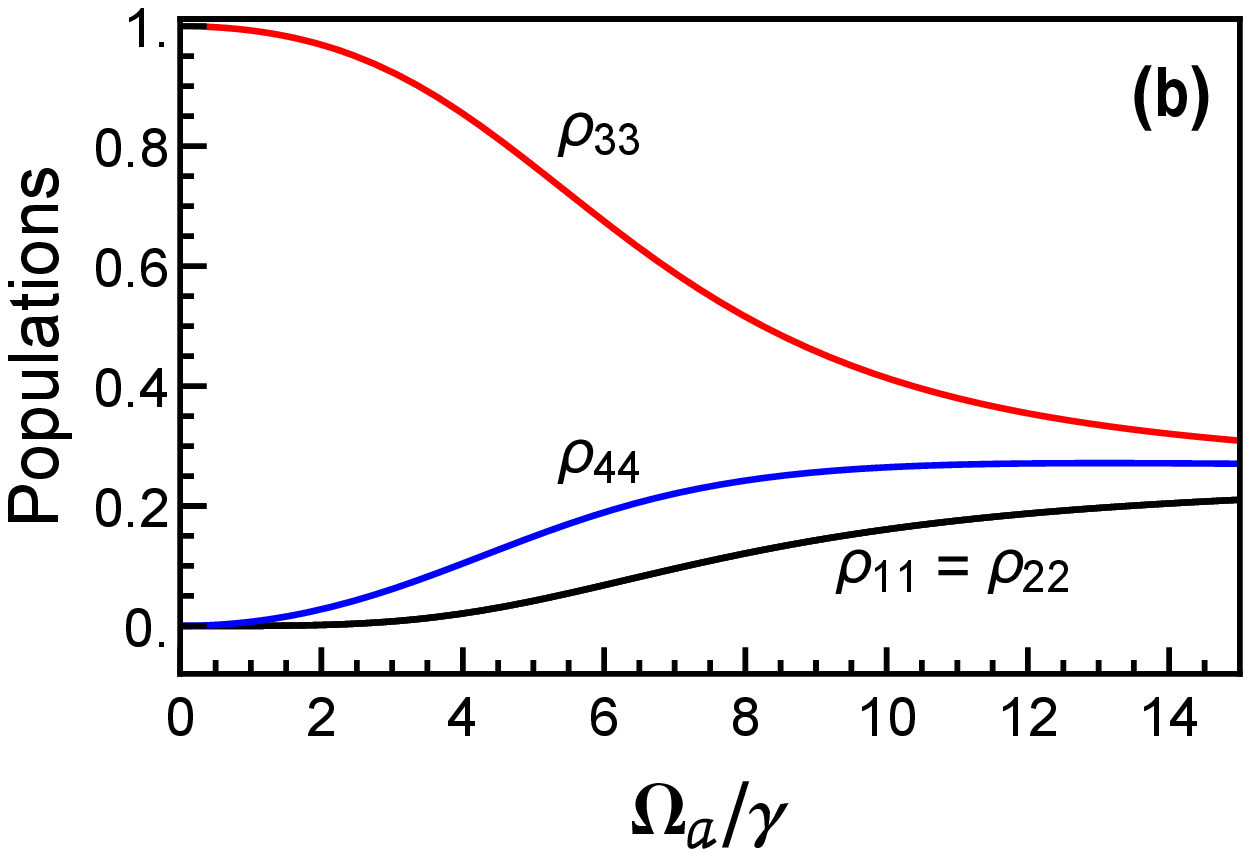}
	\end{center}
    \vspace{-1.5em}
	\caption{Steady-state results for the atomic-level populations as a function of the Rabi frequency $\Omega_a$ for
$\Delta=8\gamma $ and (a) $\Omega_{b}=0$, (b) $\Omega_{b}=12\gamma$. All the parameters are in units of $\gamma$. In all figures in
this paper, we assume that $\gamma_{12} = -\sqrt{\gamma_{1}\gamma_{2}} = -\gamma/3$ unless specified otherwise. }
\end{figure}
 \section{\label{sec:spect} RESONANCE FLUORESCENCE SPECTRUM}
 \vspace{-1em}
 We proceed to analyze the resonance fluorescence from the driven atom and derive analytic expressions suitable for numerical computation of the
 fluorescence spectra. Since the atom is driven by two coherent fields, the fluorescence fields generated by the $\pi$ and $\sigma$-transitions
 in the atom consist of coherent as well as incoherent components. We focus only on the incoherent parts of the fluorescence spectra. In the
 far-field zone, the electric field operator of the fluorescence field at an observation point $\mf{r}$ can be written as
 $\mf{\hat{E}}(\mf{r},t) = \mf{\hat{E}}^{(+)}(\mf{r},t) + \mf{\hat{E}}^{(-)}(\mf{r},t)$, where $\mf{\hat{E}}^{(+)}$
 $(\mf{\hat{E}}^{(-)}\equiv[\mf{\hat{E}}^{(+)}]^{\dag})$ represents the positive (negative) frequency part of the field. The positive frequency
 parts of the electric field operator for the fluorescence fields from the $\pi$ and $\sigma$-transitions are found to be, respectively,\cite{sbook}
\begin{flalign}
\mf{\hat{E}_{\pi}}^{\!(+)}(\mf{r},t)=&~-\frac{\omega_{o}^2}{c^2r}[\{\mf{\hat{r}}\times(\mf{\hat{r}}\times\mf{\vec{d}}_{31})\}A_{31}(t')&\nonumber\\
&+\{\mf{\hat{r}}\times (\mf{\hat{r}}\times\mf{\vec{d}}_{42})\}A_{42}(t')]e^{-i(\omega_{l}t'+\phi_{a})},&\nonumber\\
\mf{\hat{E}_{\sigma}}^{\!(+)}(\mf{r},t)=&~-\frac{\omega_{o}^2}{c^2r}[\{\mf{\hat{r}}\times(\mf{\hat{r}}\times\mf{\vec{d}}_{41})\}A_{41}(t')&\nonumber\\
&+\{\mf{\hat{r}}\times (\mf{\hat{r}}\times\mf{\vec{d}}_{32})\}&\nonumber\\
&~\times A_{32}(t')e^{-2i(\phi_{a}-\phi_{b})}] e^{-i(\omega_{l}t'+\phi_{b})},&\label{+elec}
\end{flalign}
where $t'=t-r/c$ and $\mf{\hat{r}}=\mf{r}/r$ is the unit vector along the direction of observation. By choosing the direction $(\mf{\hat{r}})$ of
detection of the fluorescence light to be along the y-direction [see Fig. 1(b)], it is seen from Eqs.(\ref{+elec}) that the fluorescence field of
the $\pi$ transitions will be polarized along $\mf{e}_{z}$ and the light emitted from the $\sigma$-transitions will be linearly polarized along
$\mf{e}_{x}$. With this choice of the detection scheme, the fluorescence light from the $\pi$ and $\sigma$-transitions can be differentiated by
means of a polarization filter.

The incoherent spectrum of resonance fluorescence is defined as
\begin{flalign}
S(\omega)=&~\frac{1}{\pi}\textit{Re}\! \int_{0}^{\infty}\! \! \!\displaystyle{\lim_{t \to \infty}} \langle\delta\mf{\hat{E}}^{(-)}(t+\tau).
\delta\mf{\hat{E}}^{(+)}(t)\rangle e^{-i\omega\tau}d\tau \!,&\label{incspec}	
\end{flalign}
where $\delta\mf{\hat{E}}^{(\pm)}=\mf{\hat{E}}^{(\pm)}\!\!-\langle\mf{\hat{E}}^{(\pm)}\!\rangle_{st}$ are the deviations of the electric field
operators $\mf{\hat{E}}^{(\pm)}$ from their steady-state average values $\langle\mf{\hat{E}}^{(\pm)}\!\rangle_{st}$. Substituting for
$\mf{\hat{E}}^{(\pm)}$ from Eq. (\ref{+elec}) into Eq. (\ref{incspec}), we get the following expressions for the incoherent spectra of the
fluorescence light emitted on the $\pi$ and $\sigma$-transitions:
\begin{flalign}
S^{\pi}(\tilde{\omega})=\frac{f_{\pi}}{\pi}\textit{Re}\! \int_{0}^{\infty}\! \! \!&\displaystyle{\lim_{t \to \infty}} [\gamma_1
\langle \delta A_{13}(t+\tau)\delta A_{31}(t)\rangle \nonumber & \\
&+ \gamma_2 \langle \delta A_{24}(t+\tau) \delta A_{42}(t) \rangle \label{spec1} & \\
&+ \gamma_{12} \langle\delta A_{13}(t+\tau) \delta A_{42}(t)\rangle \nonumber & \\
&+ \gamma_{12} \langle\delta A_{24}(t+\tau) \delta A_{31}(t)\rangle ] e^{-i\tilde{\omega}\tau}d\tau, \nonumber  &
\end{flalign}
\begin{flalign}
S^{\sigma}(\tilde{\omega})=\frac{f_{\sigma}\gamma_{\sigma}}{\pi}\textit{Re}\! &\int_{0}^{\infty}\!\displaystyle{\lim_{t \to \infty}}
[\langle \delta A_{14}(t+\tau)\delta A_{41}(t)\rangle \nonumber & \\
&+ \langle \delta A_{23}(t+\tau) \delta A_{32}(t) \rangle \label{spec2} & \\
&+ e^{-2i\phi} \langle\delta A_{14}(t+\tau) \delta A_{32}(t)\rangle \nonumber & \\
&+ e^{2i\phi} \langle\delta A_{23}(t+\tau) \delta A_{41}(t)\rangle ] e^{-i\tilde{\omega}\tau}d\tau. \nonumber  &
\end{flalign}
Here $\phi=\phi_{a}-\phi_{b}$ represents the relative phase of the applied fields, $\tilde{\omega}=\omega-\omega_{l}$ is the difference between
the frequencies of the observed radiation and the applied lasers, $\delta A_{ij}(t)=A_{ij}(t)-\langle A_{ij} \rangle_{st}$ are the fluctuations
of the atomic operators about their steady-state mean values, and $f_{\pi}$ and $f_{\sigma}$ are common prefactors. In what follows, we set the
prefactors $f_{\pi}$, $f_{\sigma}$ to unity for convenience.

To evaluate the correlation functions in Eqs. (\ref{spec1}) and (\ref{spec2}), we define a column vector of two-time correlation functions given by
\begin{align}
\hat{U}^{mn}(t,\tau)&=& \nonumber \\
&\left[\langle \delta A_{11}(t+\tau)\delta A_{mn}(t)\rangle,\langle \delta A_{33}(t+\tau)\delta A_{mn}(t)\rangle,\right.&\nonumber
\\&\left.\langle \delta A_{44}(t+\tau)\delta A_{mn}(t)\rangle,\langle \delta A_{12}(t+\tau)\delta A_{mn}(t)\rangle,\right.&\nonumber\\
&\left.\langle \delta A_{21}(t+\tau)\delta A_{mn}(t)\rangle,\langle \delta A_{13}(t+\tau)\delta A_{mn}(t)\rangle,\right.&\nonumber\\
&\left.\langle \delta A_{31}(t+\tau)\delta A_{mn}(t)\rangle,\langle \delta A_{23}(t+\tau)\delta A_{mn}(t)\rangle,\right.&\nonumber\\
&\left.\langle \delta A_{32}(t+\tau)\delta A_{mn}(t)\rangle,\langle \delta A_{14}(t+\tau)\delta A_{mn}(t)\rangle,\right.&\nonumber\\
&\left.\langle \delta A_{41}(t+\tau)\delta A_{mn}(t)\rangle,\langle \delta A_{24}(t+\tau)\delta A_{mn}(t)\rangle,\right.&\nonumber\\
&\left.\langle \delta A_{42}(t+\tau)\delta A_{mn}(t)\rangle,\langle \delta A_{34}(t+\tau)\delta A_{mn}(t)\rangle,\right.&\nonumber\\
&\left.\langle \delta A_{43}(t+\tau)\delta A_{mn}(t)\rangle\right]^{T}\!\!\!. \label{2tim}&
\end{align}
According to the quantum regression theorem \cite{lax}, the column vector (\ref{2tim}) satisfies the following equation
\begin{equation}
\frac{d}{d\tau}\hat{U}^{mn}(t,\tau)=\hat{M}\hat{U}^{mn}(t,\tau),\label{2timdiff}
\end{equation}
where the matrix $\hat{M}$ is defined as in Eq. (\ref{stdyeq}). Solving the above equation and substituting the results for the correlation
functions in Eqs. (\ref{spec1}) and (\ref{spec2}), the incoherent fluorescence spectra can be obtained as
\begin{flalign}
S^{\pi}(\tilde{\omega})=&\frac{\gamma}{3\pi}\textit{Re}\bigg\{\sum_{j=1}^{15}\displaystyle{\lim_{t \to \infty}}[\hat{N}_{6,j}
\hat{U}_{j}^{31}(t,0)\nonumber& \\
&~~~~+\hat{N}_{12,j}\hat{U}_{j}^{42}(t,0)-\hat{N}_{6,j}\hat{U}_{j}^{42}(t,0)\nonumber \\
&~~~~-\hat{N}_{12,j}\hat{U}_{j}^{31}(t,0)]\bigg\},&\label{piinc}
\end{flalign}
\begin{flalign}
S^{\sigma}(\tilde{\omega})=&\frac{2\gamma}{3\pi}\textit{Re}\bigg\{\sum_{j=1}^{15}\displaystyle{\lim_{t \to \infty}} [\hat{N}_{10,j}
\hat{U}_{j}^{41}(t,0)\nonumber  &\\
&~~~~+\hat{N}_{8,j}\hat{U}_{j}^{32}(t,0)+e^{-2i\phi}\hat{N}_{10,j}\hat{U}_{j}^{32}(t,0)\nonumber & \\
&~~~~+e^{2i\phi}\hat{N}_{8,j}\hat{U}_{j}^{41}(t,0)]\bigg\},&\label{siginc}
\end{flalign}
where $\hat{N}_{i,j}$ is the $(i,j)$ element of the matrix $\hat{N}=(i\tilde{\omega}\hat{I}-\hat{M})^{-1}$ with $\hat{I}$ being the $15 \times 15$
identity matrix.
\vspace{-1.5em}
\section{\label{sec:num}NUMERICAL RESULTS AND DRESSED-STATE ANALYSIS}
\vspace{-1.5em}
In this section we present numerical results of the fluorescence spectra and then provide an understanding of the role of VIC using dressed-state
descriptions. The detection scheme mentioned in the previous section enables us to study the fluorescence spectrum of $\pi$ and $\sigma$-transitions
separately. In the numerical calculations, the spectra $S^{ \pi}(\tilde{\omega})$ and $S^{\sigma}(\tilde{\omega})$ are obtained using
Eqs. (\ref{piinc}) and (\ref{siginc}). All the parameters such as Rabi frequencies, detuning, and the decay rates are scaled by the total decay rate
$\gamma$.
\begin{figure}[t]
\begin{center}
\includegraphics[width=6.2cm,height=4.2cm]{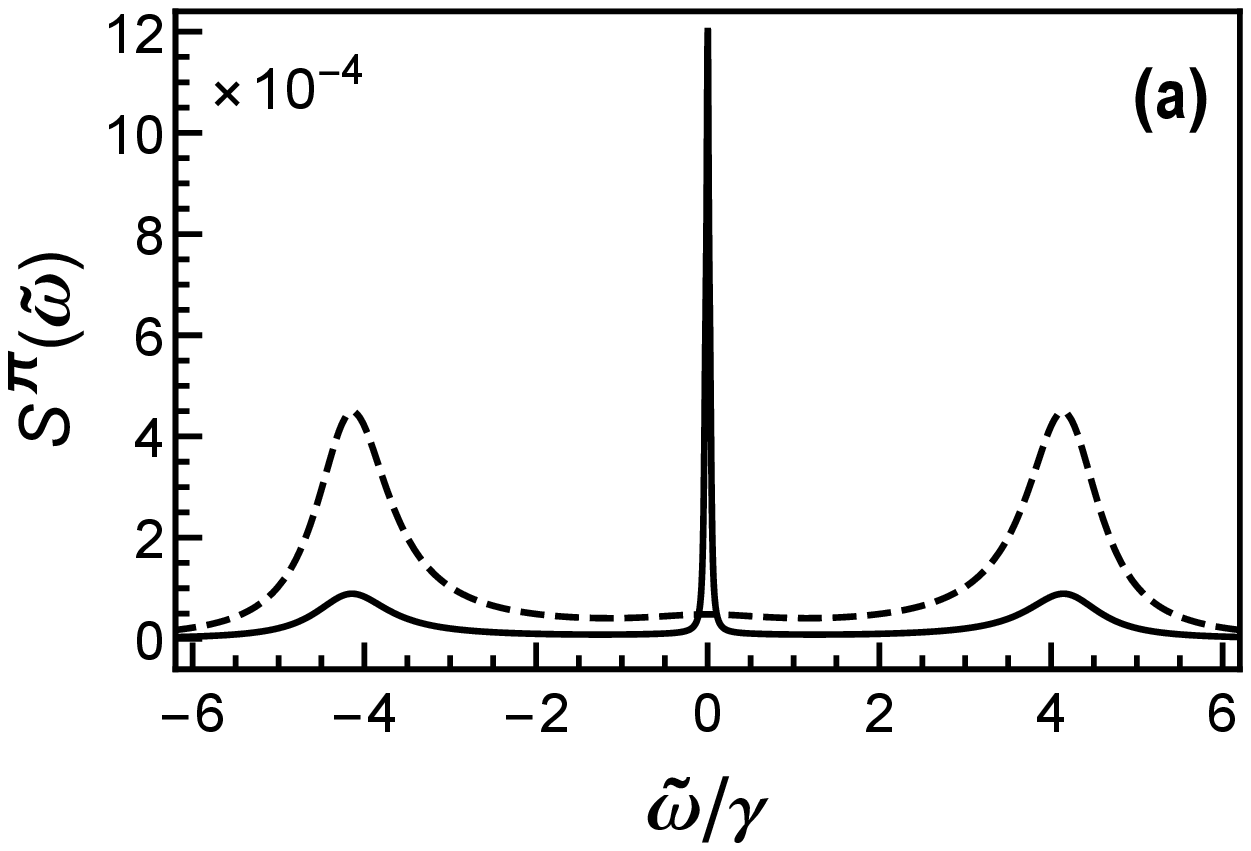}
\end{center}
\begin{center}
\includegraphics[width=6.2cm,height=4.2cm]{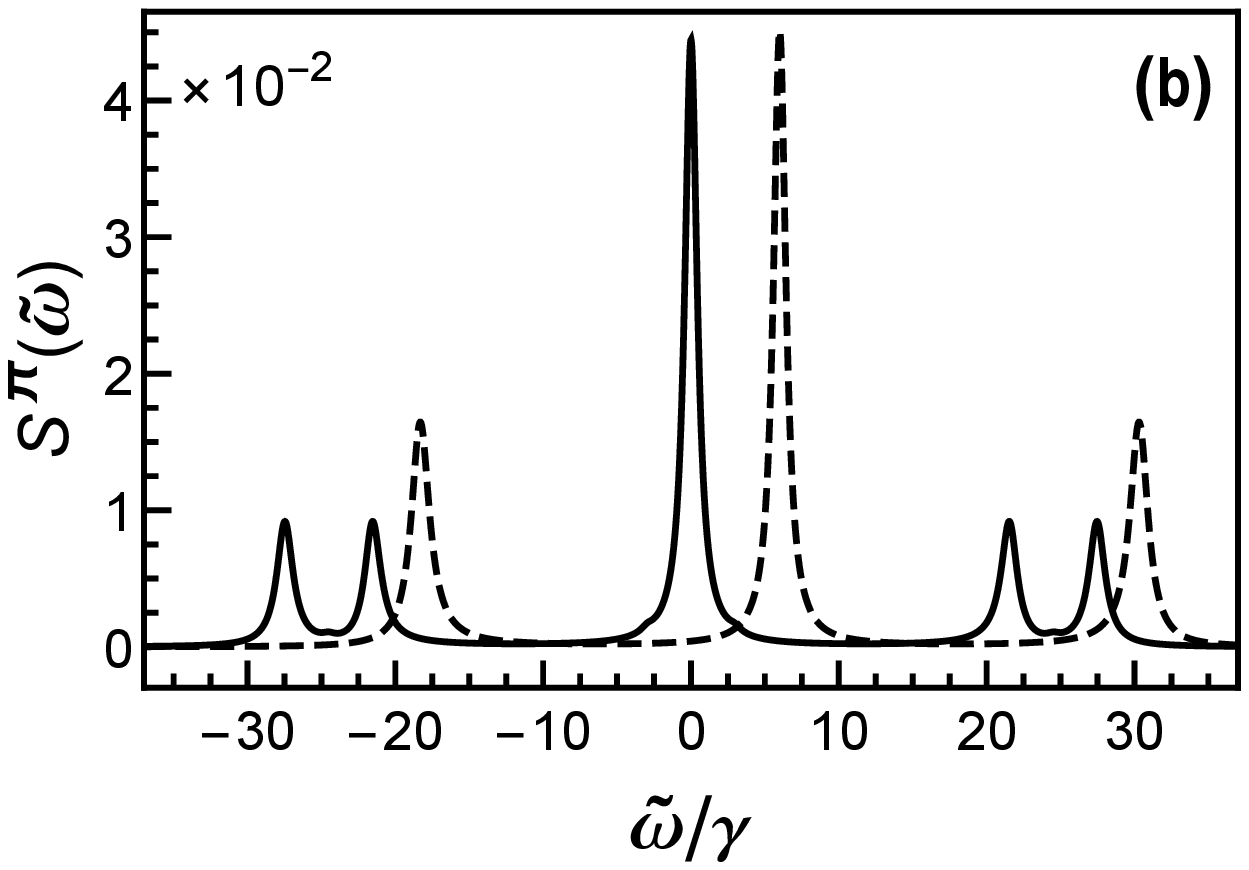}
\end{center}
\vspace{-1.5em}
\caption{The incoherent spectrum $S^{\pi}(\tilde{\omega})$ of resonance fluorescence for $\gamma=1$, $\Delta=4$, and
(a) $\Omega_{a}=0.6$, $\Omega_{b}=0.1$ and (b) $\Omega_{a}=12$, $\Omega_{b}=3$. The dashed curves represent the spectrum
for $\Omega_{b}=0$ with the remaining parameters same. Actual values of the dashed curve in (a) are 0.2 times that shown.
For clarity, the dashed curve in (b) has been shifted by 6 units along the $\tilde{\omega}/\gamma$-axis.}
\vspace{-1.5em}
\end{figure}
\subsection{Resonance fluorescence spectrum - $\pi$ transitions}
\vspace{-1.5em}
We first consider the spectrum of resonance fluorescence emitted on the $\pi$ transitions. The numerical results for the spectra are presented for
both weak and strong-driving field limits in Fig. 3. For comparison, the results are also displayed without considering the additional
$\sigma^{-}$-polarized field $(\Omega_{b}= 0)$ driving the atom. In the absence of $\sigma^{-}$-polarized field, the spectra are the same as those of
Kiffner \textit{et al.} \cite{kif1, kif2} as shown by the dashed curves in Fig. 3. It is seen that the spectrum changes significantly when the additional
field $(\Omega_{b}\neq 0)$ drives the atom. A sharp spectral peak with a width smaller than the decay rate $\gamma$ can be seen at the laser frequency
in the presence of weak driving fields [solid curve in Fig. 3(a)]. For stronger excitations ($\Omega_{a},\Omega_{b}\gg\gamma$), the additional field may
cause splitting of the sidebands in the spectrum [compare solid and dashed curves in Fig. 3(b)].

\begin{figure}[b]
	\begin{center}
     \includegraphics[width=6.3cm,height=7.3cm]{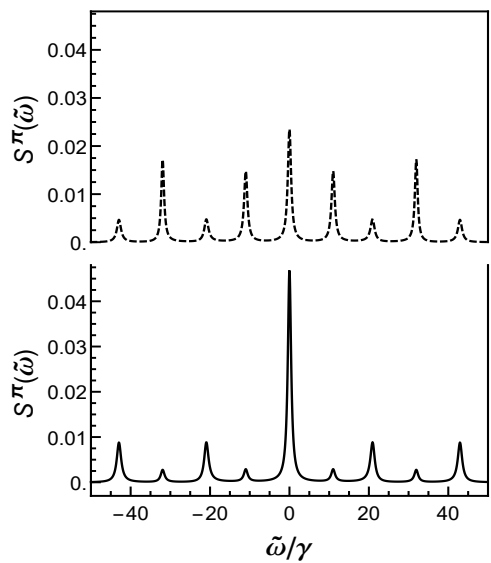}
	\end{center}
    \vspace{-1.5em}
	\caption{\small{The incoherent spectrum $S^{\pi}(\tilde{\omega})$ of resonance fluorescence for $\gamma=1$, $\Delta=0$,
$\Omega_{a}=15$, $\Omega_{b}=11$, and $\gamma_{12} = -\gamma/3$ (solid curve) and $\gamma_{12}=0$ (dashed curve).
The solid (dashed) curve represents the spectrum with (without) the VIC-terms.} \vspace{-1em}}
\end{figure}
To study the role of VIC in the fluorescence of $\pi$ transitions, we compare the spectra with and without VIC terms in Fig. 4. The spectrum without
considering VIC is obtained by setting $\gamma_{12}=0$ in Eqs. (\ref{stdyeq}) and (\ref{spec1}). Note that the last two terms with a minus sign in the
spectrum (\ref{piinc}) arise from $\gamma_{12}$-terms in Eq. (\ref{spec1}) and hence these terms do not contribute when the VIC effect is not considered.
The $\pi$-fluorescence spectrum in the absence of VIC is thus given by
\begin{flalign}
S^{\pi}(\tilde{\omega})= \frac{\gamma}{3\pi}\textit{Re}\bigg\{\sum_{j=1}^{15}\displaystyle{\lim_{t \to \infty}} &[\hat{N}_{6,j}\hat{U}_{j}^{31}(t,0)
\nonumber \\
&~~+\hat{N}_{12,j}\hat{U}_{j}^{42}(t,0)]\bigg\}.&\label{pinoint}
\end{flalign}
It is clear that the VIC modifies all the peaks of the fluorescence spectrum [see Fig. 4]. When VIC is included $(\gamma_{12} =
-\sqrt{\gamma_{1}\gamma_{2}}$) , the central peak is enhanced, whereas alternate sidebands are reduced or enhanced in comparison to the case
without VIC. For a suitable set of parameters, a complete cancellation is also possible for those sidebands which got reduced due to VIC,
as shown in Fig. 5 [compare solid and dashed curves in Fig. 5].

A physical understanding of these numerical results can be obtained if we employ the dressed-state description of atom-field interactions.
The dressed atomic states are defined as eigenstates $\ket{\Psi}$ of the Hamiltonian in the interaction picture (\ref{hamil}), i.e.,
$H_{I} \ket{\Psi} = \hbar \lambda_{\Psi}\ket{\Psi}$. For simplicity, we consider only the case in which the frequencies of the driving fields
are tuned to the atomic transition frequency $(\Delta = 0)$. Under this condition, the dressed states $\ket{\Psi}(\Psi=\alpha,\beta,\kappa,\mu)$
can be expanded in terms of the bare atomic states as
\begin{align}
\ket{\alpha}=&~c_{\alpha1}\ket{1}+c_{\alpha2}\ket{2}+c_{\alpha3}\ket{3}+c_{\alpha4}\ket{4},\nonumber\\
\ket{\beta}=&~c_{\beta1}\ket{1}+c_{\beta2}\ket{2}+c_{\beta3}\ket{3}+c_{\beta4}\ket{4},\label{dress}\\
\ket{\kappa}=&~c_{\kappa1}\ket{1}+c_{\kappa2}\ket{2}+c_{\kappa3}\ket{3}+c_{\kappa4}\ket{4},\nonumber\\
\ket{\mu}=&~c_{\mu1}\ket{1}+c_{\mu2}\ket{2}+c_{\mu3}\ket{3}+c_{\mu4}\ket{4},\nonumber
\end{align}
and the eigenvalues of these dressed states are, respectively,
\begin{flalign}
\lambda_{\alpha}&=-\frac{\Omega_{1}}{2} ,& \lambda_{\beta}&=-\frac{\Omega_{2}}{2} ,& \lambda_{\kappa}&=\frac{\Omega_{2}}{2} ,
& \lambda_{\mu}&=\frac{\Omega_{1}}{2},
\end{flalign}
where $\Omega_{1}=\sqrt{4\Omega_{a}^2+\Omega_{b}^2}+\Omega_{b}$ and $\Omega_{2}=\sqrt{4\Omega_{a}^2+\Omega_{b}^2}-\Omega_{b}$ are the
effective Rabi frequencies of the driving fields. The expansion coefficients $c_{ij}$ in Eq. (\ref{dress}) are given in Appendix A. The central peak and
the different sidebands that appear in the fluorescence spectrum can be explained in terms of transitions between the dressed states
$\ket{i}\rightarrow\ket{j}$ ($i,j=\alpha,\beta,\kappa,\mu$). The peaks in the spectrum are centered at the frequencies
$\lambda_{ij}=\lambda_{i}-\lambda_{j}$ due to the dressed-state transition $\ket{i}\rightarrow\ket{j}$ ($i,j=\alpha,\beta,\kappa,\mu$). The central peak
at $\tilde{\omega}=0$ is due to transitions that occur between adjacent manifolds of the same dressed-states. The peaks at
$\pm\Omega_{1}$ and $\pm\Omega_{2}$ come from the transitions $\ket{\mu}\leftrightarrow\ket{\alpha}$ and $\ket{\kappa}\leftrightarrow\ket{\beta}$,
respectively, whereas the transitions $\ket{\mu}\leftrightarrow\ket{\beta}$ and $\ket{\kappa}\leftrightarrow\ket{\alpha}$ are coupled to each other
and contribute to the outer sidebands located at $\pm(\Omega_{1}+\Omega_{2})/2$. The innermost sidebands at $\pm(\Omega_{1}-\Omega_{2})/2$ originate
from the coupled dressed-state transitions $\ket{\mu}\leftrightarrow\ket{\kappa}$ and $\ket{\beta}\leftrightarrow\ket{\alpha}$.

The dressed atomic states (\ref{dress}) can be used as basis states to solve for the atomic dynamics including the spontaneous decay processes. To this end,
we first rewrite the density matrix equations using the states (\ref{dress}). When the driving fields are intense $(\Omega_{a},\Omega_{b}\gg\gamma)$,
it is appropriate to invoke the secular approximation in which the couplings between populations and coherences can be ignored in the dressed-state
basis. Following closely the procedure as in the work of Narducci {\it et al.} \cite{nard}, we recast the equations of motion of the density matrix
elements (\ref{rho1})-(\ref{rho9}) using the basis of dressed atomic states (\ref{dress}) as
\begin{figure}[t]
	\begin{center}
    \includegraphics[width=6.2cm,height=4.2cm]{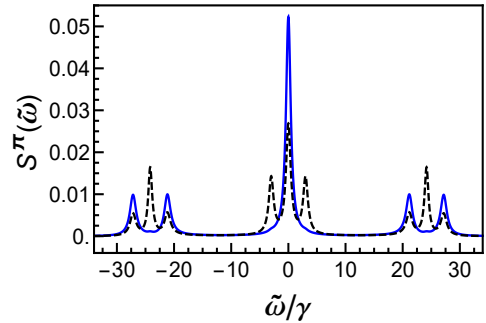}
	\end{center}
    \vspace{-1.5em}
	\caption{The incoherent spectrum $S^{\pi}(\tilde{\omega})$ of resonance fluorescence for $\gamma=1$, $\Delta=0$,
$\Omega_{a}=12$, $\Omega_{b}=3$, and $\gamma_{12} = -\gamma/3$ (solid curve) and $\gamma_{12}=0$ (dashed curve).
The solid (dashed) curve represents the spectrum with (without) the VIC-terms.}
 \end{figure}
\begin{equation}
\dot{\rho}_{\alpha\alpha}=-\Gamma_{0}\rho_{\alpha\alpha}+\Gamma \rho_{\beta\beta}+\Gamma \rho_{\kappa\kappa}+
\tilde{\Gamma}\rho_{\mu\mu}, \nonumber
\end{equation}
\begin{equation}
\dot{\rho}_{\beta\beta}= -\Gamma_{0}\rho_{\beta\beta}+\Gamma \rho_{\alpha\alpha}+ \tilde{\Gamma} \rho_{\kappa\kappa}+
\Gamma \rho_{\mu\mu}, \nonumber
\end{equation}
\begin{equation}
\dot{\rho}_{\kappa\kappa}= -\Gamma_{0}\rho_{\kappa\kappa}+\Gamma \rho_{\alpha\alpha}+ \tilde{\Gamma}\rho_{\beta\beta}
+\Gamma \rho_{\mu\mu}, \nonumber
\end{equation}
\begin{equation}
\dot{\rho}_{\mu\mu}= -\Gamma_{0}\rho_{\mu\mu} + \tilde{\Gamma} \rho_{\alpha\alpha}+\Gamma \rho_{\beta\beta} +
\Gamma \rho_{\kappa\kappa}, \nonumber
\end{equation}
\begin{equation}
\dot{\rho}_{\mu\alpha}= -(\Gamma_{1}+i\lambda_{\mu\alpha})\rho_{\mu\alpha}, \label{dreseq}
\end{equation}
\begin{equation}
\dot{\rho}_{\kappa\beta}= -(\Gamma_{2}+i\lambda_{\kappa\beta})\rho_{\kappa\beta}, \nonumber
\end{equation}
\begin{equation}
\dot{\rho}_{\mu\beta}= -(\Gamma_{3}+i\lambda_{\mu\beta})\rho_{\mu\beta}+\Gamma_{4}\rho_{\kappa\alpha},
\nonumber
\end{equation}
\begin{equation}
\dot{\rho}_{\kappa\alpha}= -(\Gamma_{3}+i\lambda_{\kappa\alpha})\rho_{\kappa\alpha}+\Gamma_{4}\rho_{\mu\beta},
\nonumber
\end{equation}
\begin{equation}
\dot{\rho}_{\mu\kappa}= -(\Gamma_{5}+i\lambda_{\mu\kappa})\rho_{\mu\kappa}-\Gamma_{6}\rho_{\beta\alpha},
\nonumber
\end{equation}
\begin{equation}
\dot{\rho}_{\beta\alpha}= -(\Gamma_{5}+i\lambda_{\beta\alpha})\rho_{\beta\alpha}-\Gamma_{6}\rho_{\mu\kappa},
\nonumber
\end{equation}
with
\begin{equation}
\Gamma_{0}=\frac{\gamma\left(9\Omega_a^2+2\Omega_b^2\right)+3\gamma_{12}\Omega_a^2}{6(4\Omega_{a}^2+\Omega_{b}^2)},
\nonumber
\end{equation}
\begin{equation}
\Gamma =\frac{\gamma\left(6\Omega_a^2+\Omega_b^2\right)+6\gamma_{12}\Omega_a^2}{12(4\Omega_{a}^2+\Omega_{b}^2)},
\nonumber
\end{equation}
\begin{equation}
\tilde{\Gamma} =\frac{\gamma\left(3\Omega_a^2+\Omega_b^2\right)-3\gamma_{12}\Omega_a^2}{6(4\Omega_{a}^2+\Omega_{b}^2)},
\nonumber
\end{equation}
\begin{equation}
\Gamma_{1}=\Gamma_{2}=\frac{\gamma\left(15\Omega_a^2+4\Omega_b^2\right)-3\gamma_{12}\Omega_a^2}{6(4\Omega_{a}^2+\Omega_{b}^2)},
\label{gterm}
\end{equation}
\begin{equation}
\Gamma_{3}=\frac{\gamma\left(11\Omega_a^2+3\Omega_b^2\right)-3\gamma_{12}\Omega_a^2}{6(4\Omega_{a}^2+\Omega_{b}^2)},
\nonumber
\end{equation}
\begin{equation}
\Gamma_{4}=\Gamma_{6}=\frac{2\gamma\Omega_{a}^2-3\gamma_{12}(2\Omega_{a}^2+\Omega_{b}^2)}{12(4\Omega_{a}^2+\Omega_{b}^2)},
\nonumber
\end{equation}
\begin{equation}
\Gamma_{5}=\frac{\gamma\left(13\Omega_a^2+3\Omega_b^2\right)+3\gamma_{12}\Omega_a^2}{6(4\Omega_{a}^2+\Omega_{b}^2)}.
\nonumber
\end{equation}
In steady state, all coherences between dressed states vanish and only the populations of the dressed states are non-zero.
After solving Eqs. (\ref{dreseq}) in  the steady-state limit ($t\rightarrow\infty$), the populations of the dressed-states are found to be
\begin{align}
\rho_{\alpha\alpha}=\rho_{\beta\beta}=\rho_{\kappa\kappa}=\rho_{\mu\mu}=\frac{1}{4}. \label{stddrespop}
\end{align}

In order to understand why certain peaks are enhanced whereas other peaks are diminished by VIC in the $\pi$-fluorescence, we calculate the
spectrum (\ref{spec1}) using the dressed states (\ref{dress}). Under the secular approximation, the incoherent fluorescence spectrum can
be worked out in a compact form as (in units of $\gamma/3$)
\begin{flalign}
S^{\pi}(\tilde{\omega})=& \frac{A_{\pi1}}{\pi}\frac{\gamma/2}{\left[\tilde{\omega}^2+\gamma^2/4\right]} + \frac{A_{\pi2}}{\pi}
\frac{\Gamma_1}{\left[(\tilde{\omega}\mp\Omega_1)^2+\Gamma_1^{2}\right]} \nonumber & \\
& + \frac{A_{\pi3}}{\pi}\frac{\Gamma_2}{\left[(\tilde{\omega}\mp\Omega_2)^2+\Gamma_2^{2}\right]} \nonumber & \\
&+ A_{\pi4} \left\{ \frac{\mc{W}_1}{\pi} \frac{\Gamma_3+\Gamma_{4}}{\left[(\tilde{\omega}\mp\left(\frac{\Omega_1+\Omega_2}{2}\right))^2 +(\Gamma_3
+\Gamma_{4})^{2}\right]} \right. \nonumber & \\
&~~~~~~~~ \left. + \frac{\mc{W}_2}{\pi} \frac{\Gamma_3-\Gamma_{4}}{\left[(\tilde{\omega}\mp\left(\frac{\Omega_1+\Omega_2}{2}\right))^2 +(\Gamma_3
-\Gamma_{4})^{2}\right]} \right\}\nonumber & \\
& + A_{\pi5} \left\{ \frac{\mc{W}_1}{\pi} \frac{\Gamma_5+\Gamma_{6}}{\left[(\tilde{\omega}\mp\Omega_b)^2+(\Gamma_5+\Gamma_{6})^{2}\right]},
\right. \label{dspect1} & \\
&~~~~~~~~~ \left. + \frac{\mc{W}_2}{\pi} \frac{\Gamma_5-\Gamma_{6}}{\left[(\tilde{\omega}\mp\Omega_b)^2+(\Gamma_5-\Gamma_{6})^{2}\right]}
\right\} \nonumber &
\end{flalign}
where
\begin{eqnarray}
A_{\pi1}&=& \frac{(\gamma-3\gamma_{12})\Omega_{a}^2}{6(4\Omega_{a}^2+\Omega_{b}^2)}, \nonumber\\
A_{\pi2}&=& A_{\pi3}=\frac{(\gamma-3\gamma_{12})\Omega_{a}^2}{24(4\Omega_{a}^2+\Omega_{b}^2)}, \nonumber \\ A_{\pi4}&=&A_{\pi5}=\frac{\gamma(2\Omega_{a}^2+\Omega_{b}^2)+6\gamma_{12}\Omega_{a}^2}{24(4\Omega_{a}^2+\Omega_{b}^2)}, \label{weight1} \\
\mc{W}_1 &=& \frac{(\gamma-3\gamma_{12})\Omega_{b}^2}{4(\gamma+3\gamma_{12})\Omega_{a}^2+2\gamma \Omega_{b}^2},\nonumber\\
\mc{W}_2 &=& \frac{(\gamma+3\gamma_{12})(4\Omega_{a}^2+\Omega_{b}^2)}{4(\gamma+3\gamma_{12})\Omega_{a}^2+2\gamma \Omega_{b}^2}.
\nonumber
\end{eqnarray}
The upper and lower signs in Eq. (\ref{dspect1}) give the positive $(\tilde{\omega}>0)$ and negative $(\tilde{\omega}<0)$ parts of
the spectrum along the $\tilde{\omega}$-axis, respectively. It is seen that the spectrum (\ref{dspect1}) is composed of nine spectral
curves, consistent with the results in Fig. 4, with widths depending upon the decay rates $(\Gamma ~\hbox{terms})$ of the dressed-state
coherences. In the absence of VIC $(\gamma_{12}=0)$, the spectral components, located at $\tilde{\omega}=\pm(\Omega_1+\Omega_2)/2$ and
$\tilde{\omega}=\pm\Omega_b$, consist of a sum of two Lorentzians with widths $2(\Gamma_3\pm\Gamma_{4})$ and $2(\Gamma_5\pm\Gamma_{6})$,
respectively [see the terms inside the curly brackets in Eq. (\ref{dspect1})]. The weights $\mc{W}_1$ and $\mc{W}_2$ of these two Lorentzians
are normalized so that $\mc{W}_1+\mc{W}_2=1$. When the effects of VIC are considered $(\gamma_{12}=-\gamma/3)$, then $\mc{W}_1=1$,
$\mc{W}_2=0$, each spectral curve is represented by a single Lorentzian as seen from Eqs. (\ref{dspect1}) and (\ref{weight1}).

The weight of each spectral curve $(A_{\pi k}, k=1,2,3,4,5)$ in the spectrum (\ref{dspect1}) can be calculated by knowing the transition rates
between the dressed states. The rate for dressed-state transition $\ket{i}\rightarrow\ket{j}$ ($i,j=\alpha,\beta,\kappa,\mu$) is given by
$\Gamma_{ij}^{\pi} = |\bra{j}P^{+}_{\pi}\ket{i}|^{2}$ where the atomic polarization operator $P^{+}_{\pi}$ for the $\pi$-fluorescence
is defined in the interaction picture as
\begin{equation}
P^{+}_{\pi} = [\sqrt{\gamma_1} A_{31} - \sqrt{\gamma_2} A_{42}] e^{-i(\omega_{l}t+\phi_{a})}. \label{pidipole}
\end{equation}
In general, the weight of the Lorentzian line originating from a single dressed-state transition $\ket{i}\rightarrow\ket{j}$ can be obtained
in the dressed-state picture as \cite{cohen}
\begin{align}
\mc{W}_{ij}^\pi=\Gamma_{ij}^{\pi}\rho_{ii},
\label{wei}
\end{align}
where $\rho_{ii}$ denotes the steady-state population of the dressed state $\ket{i}$. The derivation of the weights of the spectral
peaks (\ref{weight1}) using Eq. (\ref{wei}) is outlined in Appendix B. Since the transition rates satisfy
$\Gamma_{ij}^{\pi} = \Gamma_{ji}^{\pi}$ and all the dressed-state populations are equal in steady state [see Eq. (\ref{stddrespop})],
the spectrum $S^{\pi}(\tilde{\omega})$ in Eq. (\ref{dspect1}) is symmetric about $\tilde{\omega}=0$, confirming the
numerical results in Figs. 4 and 5. It is clear that VIC modifies all the fluorescence peaks through $\gamma_{12}$ terms in
Eq. (\ref{weight1}). As seen from  Eqs. (\ref{dspect1}) and (\ref{weight1}), the central peak and those of the sidebands at
$\lambda_{\mu\alpha}$, $\lambda_{\kappa\beta}$ are enhanced due to VIC (note that $\gamma_{12}$ is negative), whereas the peaks
located at $\lambda_{\mu\beta} (=\lambda_{\kappa\alpha})$ and $\lambda_{\mu\kappa} (=\lambda_{\beta\alpha})$ get diminished on
comparing the case with $\gamma_{12}=0$ [see Fig. 4].
\begin{figure}[t]
\begin{center}
\includegraphics[width=6.2cm,height=4.2cm]{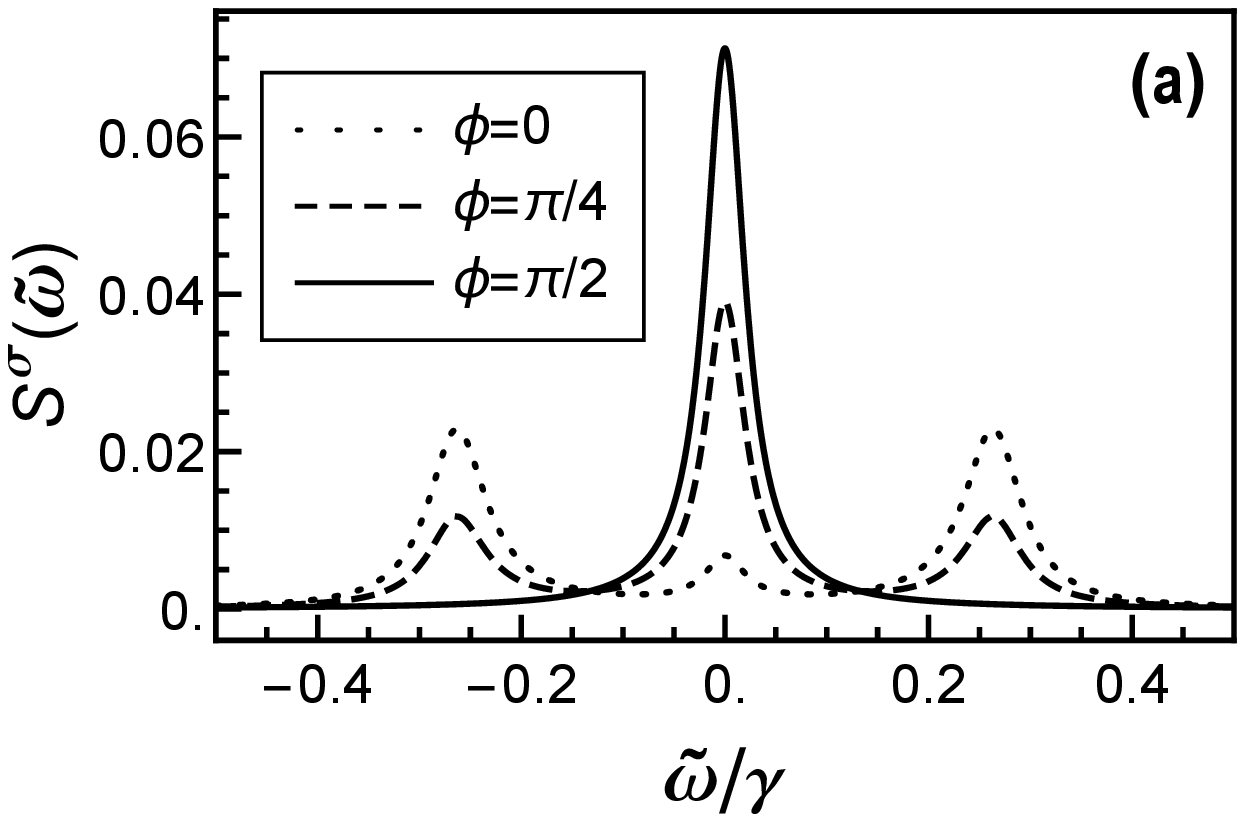}
\end{center}
\begin{center}
\includegraphics[width=6.2cm,height=4.2cm]{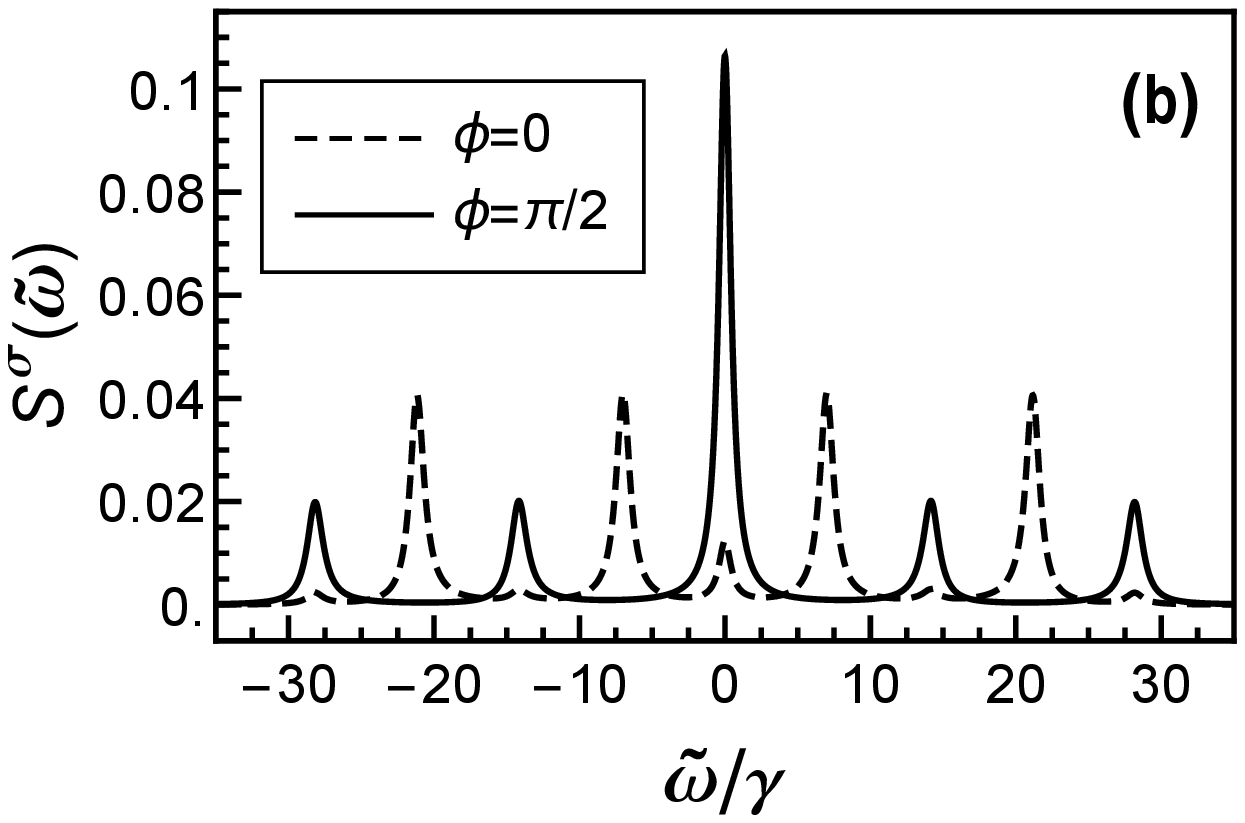}
\end{center}
\vspace{-2em}
\caption{The incoherent spectrum $S^{\sigma}(\tilde{\omega})$ of resonance fluorescence for various relative phases $\phi$.
The parameters are (a) $\Delta=4$, $\Omega_{a}=0.6$, $\Omega_{b}=0.8$ and (b) $\Delta=0$, $\Omega_{a}=10$, $\Omega_{b}=7$
with $\gamma=1$.}
\vspace{-0.8em}
\end{figure}
\vspace{-1em}
\subsection{Resonance fluorescence spectrum - $\sigma$ transitions}
\vspace{-1em}
We now investigate the spectrum of the fluorescence light emitted on the $\sigma$ transitions. When there is no
additional field $(\Omega_b = 0)$, the cross-correlation terms $\langle\delta A_{14}(t+\tau) \delta A_{32}(t)\rangle$ and
$\langle\delta A_{23}(t+\tau) \delta A_{41}(t)\rangle$ in Eq. (\ref{spec2}) turn out to be zero as reported by Kiffner
{\it et al.} \cite{kif2}. However, in the presence of the additional field $(\Omega_b \neq 0)$, these cross terms are non-zero
and the spectrum (\ref{spec2}) becomes dependent on the relative phase $\phi$ of the applied fields. In Fig. 6, we first show how
the spectral profile can be controlled by changing this relative phase. In the case of weak driving fields
$(\Omega_{a},\Omega_{b}<\gamma)$, there are three peaks when both fields are in phase ($\phi=0$). As we change the relative phase
to $\phi=\pi/4$, it is seen that the central peak is enhanced and the sidebands get reduced. For relative phase $\phi=\pi/2$, a complete
elimination of the sidebands occurs with a corresponding enhancement in the central peak [compare the graphs in Fig. 6(a)]. In other words,
the incoherent emissions at frequencies other than the laser frequencies can be cancelled by changing the relative phase of weak driving
fields. For strong-field excitation $(\Omega_{a},\Omega_{b}\gg\gamma)$, it can be shown that the central peak can be enhanced along with
alternate sidebands in the spectrum by adjusting the relative phase. This feature is illustrated in Fig. 6(b) where we compare
the spectra in the high-field limit for the relative phases $\phi=0$ and $\phi=\pi/2$.

To explore the reasons for the phase control of spectral features in the $\sigma$-fluorescence, an analytical formula for the fluorescence
spectrum (\ref{spec2}) is obtained in the dressed-state formalism as (in units of $2\gamma/3$)
\begin{flalign}
S^{\sigma}(\tilde{\omega})=& \frac{A_{\sigma1}}{\pi}\frac{\gamma/2}{\left[\tilde{\omega}^2+\gamma^2/4\right]} + \frac{A_{\sigma2}}{\pi}
\frac{\Gamma_1}{\left[(\tilde{\omega}\mp\Omega_1)^2+\Gamma_1^{2}\right]} \nonumber & \\
& + \frac{A_{\sigma3}}{\pi}\frac{\Gamma_2}{\left[(\tilde{\omega}\mp\Omega_2)^2+\Gamma_2^{2}\right]} \nonumber & \\
&+ \frac{A_{\sigma4}}{\pi} \frac{\Gamma_3+\Gamma_{4}}{\left[(\tilde{\omega}\mp\left(\frac{\Omega_1+\Omega_2}{2}\right))^2 +(\Gamma_3
+\Gamma_{4})^{2}\right]} \label{dspect2} & \\
& + \frac{A_{\sigma5}}{\pi}\frac{\Gamma_5+\Gamma_{6}}{\left[(\tilde{\omega}\mp\Omega_b)^2+(\Gamma_5+\Gamma_{6})^{2}\right]}, \nonumber &
\end{flalign}
with
\begin{eqnarray}
A_{\sigma1}&=& \frac{\gamma_{\sigma}}{4} \left(\frac{4\Omega_{a}^2\sin^2\phi+\Omega_{b}^2}{4\Omega_{a}^2+\Omega_{b}^2}\right),
\nonumber\\
A_{\sigma2}&=& A_{\sigma3} = \frac{\gamma_{\sigma}}{4} \left(\frac{4\Omega_{a}^2\sin^2\phi+\Omega_{b}^2}{4(4\Omega_{a}^2+\Omega_{b}^2)}\right),
\label{weight2}\\
A_{\sigma4}&=& A_{\sigma5}= \frac{\gamma_{\sigma}}{4} \left(\frac{2\Omega_{a}^2\cos^2\phi}{4\Omega_{a}^2+\Omega_{b}^2}\right). \nonumber
\end{eqnarray}
The various terms in the spectrum given above have their meanings as explained following Eq. (\ref{dspect1}). From the analytic expressions
(\ref{dspect2}) and (\ref{weight2}), it is evident that the central peak and the sidebands peaked at $\pm \Omega_{1}$
and $\pm \Omega_{2}$ in the spectrum get enhanced as the phase $\phi$ is changed from $0$ to $\pi/2$ [see Fig. 6(b)]. Note that the weights
of the Lorentzians (\ref{weight2}) depend only on the relative phase of the applied fields and are independent of the VIC parameter
$\gamma_{12}$. This is in agreement with the numerical calculations of the spectrum in the strong-field limit. The derivation of the
formulas (\ref{weight2}) follows from calculating the weight of the spectral line $\mc{W}_{ij}^\sigma= \Gamma_{ij}^{\sigma}\rho_{ii}$
[similar to Eq. (\ref{wei})] and is given in Appendix B. Here $\Gamma_{ij}^{\sigma} = |\bra{j}P^{+}_{\sigma}\ket{i}|^{2}$ represents the
transition rate between the dressed states with the atomic polarization operator defined for the $\sigma$-fluorescence as
\begin{equation}
P^{+}_{\sigma} = [\sqrt{\gamma_{\sigma}} A_{41} + \sqrt{\gamma_{\sigma}} A_{32}e^{-2i\phi}] e^{-i(\omega_{l}t+\phi_{b})}.
\label{sigdipole}
\end{equation}

Finally, we consider the effect of VIC in the $\sigma$ transitions. In Fig. 7, the spectra $S^{\sigma}(\tilde{\omega})$ with and
without VIC are plotted for a fixed relative phase $\phi=\pi/2$. The numerical results show that the central peak in the spectrum
is not affected by VIC terms in the density matrix equations (\ref{stdyeq}) as can be verified with the analytical result
(\ref{dspect2}). However, all the sideband peaks in the spectrum get reduced by VIC (compare solid and dashed curves in Fig. 7).
The reduction is more prominent for certain sidebands, whereas the influence of VIC is comparatively lesser for the other sidebands.
This can be understood from the widths $(\Gamma ~\hbox{terms})$ of the Lorentzians in the spectrum (\ref{dspect2}). For the parameters
of Fig. 7, the widths ($\Gamma_{1}$, $\Gamma_{2})$ of the spectral curves peaked at $\pm \Omega_{1}$ and $\pm \Omega_{2}$ increase
due to VIC [see Eq. (\ref{gterm})], thereby reducing the peak heights. \vspace{-0.5cm}
\begin{figure}[t]
\begin{center}
\includegraphics[width=6.2cm,height=4.2cm]{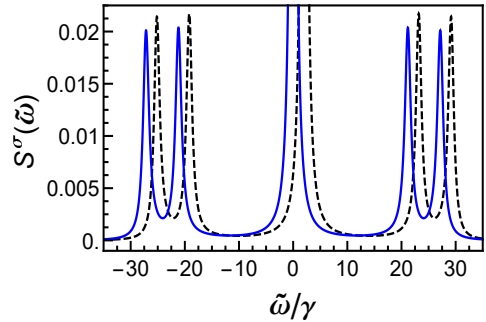}
\end{center}
\vspace{-2.2em}
\caption{The incoherent spectrum $S^{\sigma}(\tilde{\omega})$ of resonance fluorescence for $\gamma=1$, $\Delta=0$,
$\Omega_{a}=12$, $\Omega_{b}=3$, $\phi=\pi/2$, and $\gamma_{12} = -\gamma/3$ (solid curve) and $\gamma_{12}=0$ (dashed curve).
The solid (dashed) curve represents the spectrum with (without) the VIC-terms. For clarity, the dashed curve has been shifted
by 2 units along the $\tilde{\omega}/\gamma$-axis. The scale on the graphs is chosen to focus on the sidebands in the spectrum
and hence the central peaks at $\tilde{\omega}=0$ are not shown.}
\vspace{-0.5em}
\end{figure}
\section{\label{sec:concl} conclusions}
\vspace{-1.5em}
In conclusion, we have studied theoretically the effects of VIC in the resonance fluorescence spectrum of $J=1/2$ to $J=1/2$ system
driven by two coherent fields. We have found that VIC affects all the peaks of the spectrum of $\pi$ transitions, where
certain peaks are enhanced whereas others are diminished. For a suitable choice of  parameters, certain sidebands may be eliminated
due to VIC. The enhancement and suppression of the spectral peaks can be explained using the weight of the line in which the
VIC term changes the transition rates of the dressed states. The fluorescence spectrum emitted on the $\sigma$ transitions is found
to be dependent on the relative phase of the applied fields even though the steady-state atomic populations are phase-independent.
We find that the phase-dependence of the spectral profiles occurs because the detection process involves only the $\mf{e}_{x}$-polarized
field in the fluorescence light from the $\sigma$ transitions. The results of this paper can be verified experimentally using a
single laser-cooled ${}^{198}\text{Hg}^+$ ion in optical trap as in the experiment of Eichmann {\it et al.} \cite{hgion}. Further,
it would be worthwhile to investigate the effects of VIC in photon correlations and squeezing spectra in the system studied here.
Detailed descriptions of such studies will be published elsewhere.

\vspace{-0.7cm}
\appendix
\section{COEFFICIENTS IN THE DRESSED STATES}
\vspace{-1em}
The coefficients in Eq. (\ref{dress}) are
\begin{flalign}
c_{\alpha1}&=-\frac{1}{2}\sqrt{\frac{\Omega_{1}}{\Omega_{1}-\Omega_{b}}} ,
& c_{\alpha2}&=-\frac{\Omega_{a}}{\sqrt{\Omega_{1}(\Omega_{1}-\Omega_{b})}},&\nonumber\\
c_{\kappa1}&=\frac{1}{2}\sqrt{\frac{\Omega_{2}}{\Omega_{2}+\Omega_{b}}} ,
&  c_{\kappa2}&=-\frac{\Omega_{a}}{\sqrt{\Omega_{2}(\Omega_{2}+\Omega_{b})}},&\nonumber\\
c_{\alpha4}&=-c_{\mu1}=c_{\mu4}=c_{\alpha1} ,& c_{\alpha3}&=c_{\mu2}=c_{\mu3}=-c_{\alpha2}, &\nonumber\\
c_{\beta4}&=-c_{\beta1}=c_{\kappa4}=c_{\kappa1} ,& c_{\beta2}&=c_{\beta3}=c_{\kappa3}=-c_{\kappa2}.&
\label{cval}
\end{flalign}

\vspace{-1.5em}
\section{calculation of the weights of the spectral lines}
\vspace{-1em}
The purpose of this appendix is to outline how the weights of the Lorentzians in the spectra (\ref{dspect1})
and (\ref{dspect2}) can be derived. For that we use Eq. (\ref{wei}) to find the contribution of each
dressed-state transition $\ket{i}\rightarrow\ket{j}$ ($i,j=\alpha,\beta,\kappa,\mu$) to a spectral curve peaked at
$\tilde{\omega}=\lambda_{i}-\lambda_{j}$ . The total weight of the spectral curve is obtained by adding
the contributions from all dressed-state transitions that give rise to the peak. Thus, using Eq. (\ref{wei}),
the weights are found to be
\begin{eqnarray}
A_{k1}&=& \Gamma_{\alpha\alpha}^{k} \rho_{\alpha\alpha} + \Gamma_{\beta\beta}^{k} \rho_{\beta\beta} +
\Gamma_{\kappa\kappa}^{k} \rho_{\kappa\kappa} + \Gamma_{\mu\mu}^{k} \rho_{\mu\mu}, \nonumber \\
A_{k2}&=& \Gamma_{\mu\alpha}^{k} \rho_{\mu\mu} = \Gamma_{\alpha\mu}^{k} \rho_{\alpha\alpha}, \nonumber \\
A_{k3} &=& \Gamma_{\kappa\beta}^{k} \rho_{\kappa\kappa} = \Gamma_{\beta\kappa}^{k} \rho_{\beta\beta}, \label{acoeff} \\
A_{k4} &=& \Gamma_{\mu\beta}^{k} \rho_{\mu\mu} + \Gamma_{\kappa\alpha}^{k} \rho_{\kappa\kappa}
= \Gamma_{\beta\mu}^{k} \rho_{\beta\beta} + \Gamma_{\alpha\kappa}^{k} \rho_{\alpha\alpha},  \nonumber \\
A_{k5} &=& \Gamma_{\mu\kappa}^{k} \rho_{\mu\mu} + \Gamma_{\beta\alpha}^{k} \rho_{\beta\beta}
= \Gamma_{\kappa\mu}^{k} \rho_{\kappa\kappa} + \Gamma_{\alpha\beta}^{k} \rho_{\alpha\alpha}, \nonumber
\end{eqnarray}
where $k = \pi~(\sigma)$ refers to the spectrum of the $\pi$ $(\sigma)$ transitions.

On using the polarization operator (\ref{pidipole}) in the transition rates $\Gamma_{ij}^{\pi} =
|\bra{j}P^{+}_{\pi}\ket{i}|^{2}$ for the $\pi$-fluorescence, the weights are obtained from Eqs. (\ref{acoeff}) to be
\begin{flalign}
A_{\pi1}=&~\rho_{\alpha\alpha}(c_{\alpha1}^2 c_{\alpha3}^2\gamma_{1}+c_{\alpha2}^2 c_{\alpha4}^2\gamma_{2}+2\gamma_{12}
c_{\alpha1}c_{\alpha2}c_{\alpha3}c_{\alpha4})&\nonumber\\
&+\rho_{\beta\beta}(c_{\beta1}^2 c_{\beta3}^2\gamma_{1}+c_{\beta2}^2
c_{\beta4}^2\gamma_{2}+2\gamma_{12}c_{\beta1}c_{\beta2}c_{\beta3}c_{\beta4})&\nonumber\\
&+\rho_{\kappa\kappa}(c_{\kappa1}^2 c_{\kappa3}^2\gamma_{1}+c_{\kappa2}^2 c_{\kappa4}^2\gamma_{2}+2\gamma_{12}c_{\kappa1}
c_{\kappa2}c_{\kappa3}c_{\kappa4})&\nonumber\\
&+\rho_{\mu\mu}(c_{\mu1}^2 c_{\mu3}^2\gamma_{1}+c_{\mu2}^2 c_{\mu4}^2\gamma_{2}+2\gamma_{12}
c_{\mu1}c_{\mu2}c_{\mu3}c_{\mu4}), \nonumber&
\end{flalign}
\begin{flalign}
A_{\pi2}=&~\rho_{\mu\mu}(c_{\mu1}^2c_{\alpha3}^2\gamma_{1}+c_{\mu2}^2c_{\alpha4}^2\gamma_{2}+2\gamma_{12}
c_{\mu1}c_{\mu2}c_{\alpha3}c_{\alpha4}), \nonumber&
\end{flalign}
\begin{flalign}
A_{\pi3}=&~\rho_{\kappa\kappa}(c_{\kappa1}^2c_{\beta3}^2\gamma_{1}+c_{\kappa2}^2c_{\beta4}^2\gamma_{2}+2\gamma_{12}
c_{\kappa1}c_{\kappa2} c_{\beta3}c_{\beta4}), \label{api}&
\end{flalign}
\begin{flalign}
A_{\pi4}=&~\rho_{\mu\mu}(c_{\mu1}^2c_{\beta3}^2\gamma_{1}+c_{\mu2}^2c_{\beta4}^2\gamma_{2}+2\gamma_{12}c_{\mu1}
c_{\mu2}c_{\beta3}c_{\beta4})&\nonumber\\
&+\rho_{\kappa\kappa}(c_{\kappa1}^2c_{\alpha3}^2\gamma_{1}+c_{\kappa2}^2c_{\alpha4}^2\gamma_{2}+2\gamma_{12}
c_{\kappa1}c_{\kappa2}c_{\alpha3}c_{\alpha4}), \nonumber &
\end{flalign}
\begin{flalign}
A_{\pi5}=&~\rho_{\mu\mu}(c_{\mu1}^2c_{\kappa3}^2\gamma_{1}+c_{\mu2}^2c_{\kappa4}^2\gamma_{2}+2\gamma_{12}
c_{\mu1}c_{\mu2}c_{\kappa3}c_{\kappa4})&\nonumber\\
&+\rho_{\beta\beta}(c_{\beta1}^2c_{\alpha3}^2\gamma_{1}+c_{\beta2}^2c_{\alpha4}^2\gamma_{2}+2\gamma_{12}
c_{\beta1}c_{\beta2}c_{\alpha3}c_{\alpha4}). \nonumber &
\end{flalign}

A similar calculation using the polarization operator (\ref{sigdipole}) in the transition rates
$\Gamma_{ij}^{\sigma} = |\bra{j}P^{+}_{\sigma}\ket{i}|^{2}$ for the $\sigma$-fluorescence and the
equations (\ref{acoeff}) leads to
\begin{flalign}
A_{\sigma1}=&~\rho_{\alpha\alpha}\gamma_{\sigma}(c_{\alpha1}^2 c_{\alpha4}^2
+c_{\alpha2}^2 c_{\alpha3}^2+2c_{\alpha1}c_{\alpha2}c_{\alpha3}c_{\alpha4}
\cos2\phi)&\nonumber\\
&+\rho_{\beta\beta}\gamma_{\sigma}(c_{\beta1}^2 c_{\beta4}^2+c_{\beta2}^2
c_{\beta3}^2 +2c_{\beta1} c_{\beta2}c_{\beta3}c_{\beta4}\cos2\phi)&\nonumber\\
&+\rho_{\kappa\kappa}\gamma_{\sigma}(c_{\kappa1}^2 c_{\kappa4}^2+c_{\kappa2}^2
c_{\kappa3}^2+2c_{\kappa1}c_{\kappa2}c_{\kappa3}c_{\kappa4}\cos2\phi)&\nonumber\\
&+\rho_{\mu\mu}\gamma_{\sigma}(c_{\mu1}^2 c_{\mu4}^2+c_{\mu2}^2 c_{\mu3}^2
+2c_{\mu1}c_{\mu2} c_{\mu3}c_{\mu4}\cos2\phi), \nonumber &
\end{flalign}
\begin{flalign}
A_{\sigma2}=&~\rho_{\mu\mu}\gamma_{\sigma}(c_{\mu1}^2 c_{\alpha4}^2+c_{\mu2}^2
c_{\alpha3}^2 +2c_{\mu1}c_{\mu2}c_{\alpha3}c_{\alpha4}\cos2\phi), \nonumber &
\end{flalign}
\vspace{0.5em}
\begin{flalign}
A_{\sigma3}=&~\rho_{\kappa\kappa}\gamma_{\sigma}(c_{\kappa1}^2c_{\beta4}^2+
c_{\kappa2}^2c_{\beta3}^2+2c_{\kappa1}c_{\kappa2}c_{\beta3}c_{\beta4}\cos2\phi),
\label{asig} &
\end{flalign}
\begin{flalign}
A_{\sigma4}=&~\rho_{\mu\mu}\gamma_{\sigma}(c_{\mu1}^2 c_{\beta4}^2+c_{\mu2}^2
c_{\beta3}^2+2c_{\mu1}c_{\mu2}c_{\beta3}c_{\beta4}\cos2\phi)&\nonumber\\
&+\rho_{\kappa\kappa}\gamma_{\sigma}(c_{\kappa1}^2c_{\alpha4}^2+c_{\kappa2}^2c_{\alpha3}^2
+2c_{\kappa1}c_{\kappa2}c_{\alpha3}c_{\alpha4}\cos2\phi), \nonumber &
\end{flalign}
\begin{flalign}
A_{\sigma5}=&~\rho_{\mu\mu}\gamma_{\sigma}(c_{\mu1}^2 c_{\kappa4}^2+c_{\mu2}^2
c_{\kappa3}^2+2c_{\mu1}c_{\mu2}c_{\kappa3}c_{\kappa4}\cos2\phi)&\nonumber\\
& +\rho_{\beta\beta}\gamma_{\sigma}(c_{\beta1}^2 c_{\alpha4}^2+c_{\beta2}^2 c_{\alpha3}^2
+2c_{\beta1}c_{\beta2}c_{\alpha3}c_{\alpha4}\cos2\phi). \nonumber &
\end{flalign}
Substituting the coefficients in the dressed states (\ref{cval}) into the expressions
(\ref{api}) and (\ref{asig}) and simplifying, we get the weights (\ref{weight1}) and (\ref{weight2})
in the spectra.


\vspace{-1em}
\begin{thebibliography}{56}
    \bibitem{ficek1} Z. Ficek and S. Swain, J. Mod. Opt. {\bf 49}, 3 (2002).
    \bibitem{ficek2} Z. Ficek and S. Swain, {\it Quantum Interference and Coherence} (Springer, New York, 2005).
	\bibitem{car2} D. A. Cardimona, M. G. Raymer, and C. R. Stroud, J. Phys. B {\bf 15}, 55 (1982).
    \bibitem{java} J. Javanainen, Europhys. Lett. {\bf 17}, 407 (1992).
	\bibitem{zhu1} S. Y. Zhu, R. C. F. Chan, and C. P. Lee, Phys. Rev. A {\bf 52}, 710 (1995).
	\bibitem{zho} P. Zhou and S. Swain, Phys. Rev. Lett. {\bf 77}, 3995 (1996); J. Evers, D. Bullock, and C. H. Keitel, Opt. Commun.
     {\bf 209}, 173 (2002).
	\bibitem{gong} S. Q. Gong, E. Paspalakis, and P. L. Knight, J. Mod. Opt. {\bf 45}, 2433 (1998); E. Paspalakis, S. Q. Gong, and
     P. L. Knight, Opt. Commun. {\bf 152}, 293 (1998); P. Dong and S. H. Tang, Phys. Rev. A {\bf 65}, 033816 (2002); W. H. Xu,
     J. H. Wu, and J. Y. Gao, J. Phys. B {\bf 39}, 1461 (2006).
    \bibitem{bor} D. Bortman-Arbiv, A. D. Wilson-Gordon, and H. Friedmann, Phys. Rev. A {\bf 63}, 043818 (2001).
	\bibitem{ever1} M. Macovei, J. Evers, and C. H. Keitel, Phys. Rev. Lett. {\bf 91}, 233601 (2003).
	\bibitem{gao} S. Y. Gao, F. L. Li, and S. Y. Zhu, Phys. Rev. A {\bf 66}, 043806 (2002).
	\bibitem{zhu2} S. Y. Zhu and M. O. Scully, Phys. Rev. Lett. {\bf 76}, 388 (1996).
    \bibitem{paspa} E. Paspalakis and P. L. Knight, Phys. Rev. Lett. {\bf 81}, 293 (1998).
    \bibitem{li1} F. L. Li and S. Y. Zhu, Phys. Rev. A {\bf 59}, 2330 (1999).
	\bibitem{li2} R. Arun, Phys. Lett. A {\bf 377}, 200 (2013); F. L. Li, S. Y. Gao, and S. Y. Zhu, Phys. Rev. A {\bf 67}, 063818 (2003).
    \bibitem{hou} B. P. Hou, S. J. Wang, W. L. Yu, and W. L. Sun, Phys. Rev. A {\bf 69}, 053805 (2004).
	\bibitem{arun1} R. Arun, Phys. Rev. A {\bf 77}, 033820 (2008).
    \bibitem{yan} X. A. Yan, L. Q. Wang, B. Y. Yin, and J. P. Song, J. Opt. Soc. Am. B {\bf 26}, 1862 (2009).
	\bibitem{si} L. G. Si, X. Y. L\"u, X. Hao, and J. H. Li, J. Phys. B {\bf 43}, 065403 (2010).
    \bibitem{arun2} R. Arun, Phys. Rev. A {\bf 94}, 043843 (2016).
    \bibitem{bypass} A. J. Li, X. L. Song, X. G. Wei, L. Wang, and J. Y. Gao, Phys. Rev. A {\bf 77}, 053806 (2008);
    Z. Ficek and S. Swain, {\it ibid.} {\bf 69}, 023401 (2004); P. Zhou and S. Swain, Opt. Commun. {\bf 179}, 267 (2000);
    E. Paspalakis, C. H. Keitel, and P. L. Knight, Phys. Rev. A {\bf 58}, 4868 (1998);
	\bibitem{kif1} M. Kiffner, J. Evers, and C. H. Keitel, Phys. Rev. Lett. {\bf 96}, 100403 (2006).
	\bibitem{kif2} M. Kiffner, J. Evers, and C. H. Keitel, Phys. Rev. A {\bf 73}, 063814 (2006).
    \bibitem{polder} D. Polder and M. F. H. Schuurmans, Phys. Rev. A {\bf 14}, 1468 (1976).
    \bibitem{hgion} U. Eichmann, J. C. Bergquist, J. J. Bollinger, J. M. Gilligan, W. M. Itano, D. J. Wineland, and M. G. Raizen,
    Phys. Rev. Lett. {\bf 70}, 2359 (1993).
	\bibitem{gs4} S. Das and G. S. Agarwal, Phys. Rev. A {\bf 77}, 033850 (2008).
	\bibitem{tan} H. T. Tan, H. X. Xia, and G. X.  Li, J. Phys. B {\bf 42}, 125502 (2009).
	\bibitem{ever2} S.I.Schmid and J.Evers, Phys.Rev.A {\bf 81}, 063805 (2010).
	\bibitem{sak} J. J. Sakurai, {\it Modern Quantum Mechanics}, (Addison-Wesley, Reading, MA, 1994).
    \bibitem{sbook} M. O. Scully and M. S. Zubairy, {\it Quantum Optics} (Cambridge University Press, London, 1997).
	\bibitem{lax} M. Lax, Phys. Rev. {\bf 129}, 2342 (1963).
    \bibitem{nard} L. M. Narducci, M. O. Scully, G.-L. Oppo, P. Ru, and J. R. Tredicce, Phys. Rev. A {\bf 42}, 1630 (1990); A. S. Manka,
    H. M. Doss, L. M. Narducci, P. Ru, and G.-L. Oppo, {\it ibid.} {\bf 43}, 3748 (1991).
	\bibitem{cohen} C. Cohen-Tannoudji, J. Dupont-Roc, and G. Grynberg, {\it Atom-Photon Interactions}, (Wiley, New York, 1998).
\end{thebibliography}
\end{document}